\documentclass[twocolumn]{aastex631} 
\usepackage{amsmath}
\shorttitle{Proper Motions in the COSMOS Field}
\shortauthors{Fajardo-Acosta et al.}
\graphicspath{{./}{}}

\begin{document}
\received{2021 Nov 20}; \revised{2022 Feb 24}; \accepted{2022 Mar 1, ApJ, in press}
\title{Joint Survey Processing II: Stellar Proper Motions in the COSMOS Field, from Hubble Space Telescope ACS and Subaru Telescope HSC Observations}

\correspondingauthor{Sergio B. Fajardo-Acosta}
\email{fajardo@ipac.caltech.edu}

\author[0000-0001-9309-0102]{Sergio B. Fajardo-Acosta
 }
\affiliation{Caltech/IPAC, Mail Code 100-22, Pasadena, CA 91125, USA}

\author[0000-0002-9382-9832]{Andreas Faisst}
\affiliation{Caltech/IPAC, Mail Code 314-6, Pasadena, CA 91125, USA}

\author[0000-0003-4072-169X]{Carl J. Grillmair}
\affiliation{Caltech/IPAC, Mail Code 314-6, Pasadena, CA 91125, USA}

\author[0000-0001-7583-0621]{Ranga-Ram Chary}
\affiliation{Caltech/IPAC, Mail Code 314-6, Pasadena, CA 91125, USA}

\author[0000-0002-5158-243X]{Roberta Paladini}
\affiliation{Caltech/IPAC, Mail Code 100-22, Pasadena, CA 91125, USA}

\author[0000-0001-7648-4142]{Ben Rusholme}
\affiliation{Caltech/IPAC, Mail Code 100-22, Pasadena, CA 91125, USA}

\author[0000-0003-0987-5738]{Nathaniel Stickley}
\affiliation{Caltech/IPAC, Mail Code 100-22, Pasadena, CA 91125, USA}



\begin{abstract}

We analyze stellar proper motions in the COSMOS field to assess the presence of bulk motions. At bright magnitudes ($G-$band 18.5--20.76 AB), we use the proper motions of 1,010 stars in the Gaia DR2 catalog. At the faint end, we computed proper motions of 11,519 point-like objects at $i-$band magnitudes 19--25 AB using Hubble ACS and Subaru HSC, which span two epochs about 11 years apart. In order to measure these proper motions with unprecedented accuracy at faint magnitudes, we developed a foundational set of astrometric tools which will be required for Joint Survey Processing (JSP) of data from the next generation of optical/infrared surveys. The astrometric grids of Hubble ACS and Subaru HSC mosaics were corrected at the catalog level, using proper motion-propagated and parallax-corrected Gaia DR2 sources. These astrometric corrections were verified using compact extragalactic sources. Upon comparison of our measured proper motions with Gaia DR2, we estimate the uncertainties in our measurements to be $\sim$2--3 mas/yr per axis, down to 25.5 AB mag. We corrected proper motions for the mean motion of the Sun, and we find that late-type main-sequence stars predominantly in the thin disk in the COSMOS field have space velocities mainly towards the Galactic center. We detect candidate high-velocity ($\geq 220$ km/s) stars, 6 of them at $\sim$0.4-6 kpc from the Gaia sample, and 5 of them at $\sim$20 kpc from the faint star HSC and ACS sample. The sources from the faint star sample may be candidate halo members of the Sangarius stream.

\end{abstract}



\section{Introduction} \label{sec:intro}

The combined analysis of multi-wavelength, multi-mission data sets with different spatial resolutions and different time baselines, at the pixel 
level, can be a powerful way to uncover phenomenology
not discernible with a single data set. It requires
astrometry and photometry to be made concordant with high precision, which is sometimes a challenge given the role of source confusion at faint magnitudes, varying point spread functions, and different pixel scales. The {\it Joint Survey Processing} (hereafter referred to as JSP) effort will, as its name implies, undertake joint pixel level processing of data from {\it Euclid}, the {\it Nancy Grace Roman Space Telescope} (formerly known as the {\it Wide-Field Infrared Space Telescope} or {\it WFIRST}), and the {\it Vera C.\ Rubin Observatory} \citep[formerly known 
 as the {\it Large Synoptic Survey Telescope}, or {\it LSST}, an acronym now reserved for the {\it Rubin Observatory}'s Legacy Survey of Space and Time][]{LSST}. The spatial resolution of the space-platform surveys and their near-infrared bandpasses will enable one to distinguish stars from galaxies, while alleviating both source confusion and the role of dust extinction. In comparison, the multi-visit cadence of {\it LSST} and its $>$10-year time baseline, as well as its differential time baseline with the single-visit space surveys, will allow one to study the proper motions of these stars. Measuring stellar motions is a powerful way to measure the properties of tidal streams \citep[e.g.][]{KOPOSOV19} and bulk motions within the Galaxy.

In order to develop the capabilities
to measure stellar proper motions especially at brightnesses up to 5 mags fainter than {\it Gaia} \citep{GAIA-MISSION}, and to better understand the systematics floor that arises, we have undertaken joint survey processing on prototype datasets in the 1.64\,deg$^{2}$ Cosmological Evolution Survey (COSMOS)\footnote{\url{https://cosmos.astro.caltech.edu/}} field \citep{SCOVILLE07}. In particular, the COSMOS field has data over a decade in time, first from the {\it Hubble Space Telescope} ({\it HST}) Advanced Camera for Surveys \citep[ACS,][]{KOEKEMOER07}\footnote{\url{https://www.stsci.edu/hst/instrumentation/acs}} and more recently, the {\it Subaru Telescope Hyper Suprime-Cam} camera \citep[HSC,][]{MIYASAKI2018,AIHARA18}\footnote{\url{https://www.naoj.org/Projects/HSC/}} with spatial resolution of $\sim0.1\arcsec$ and $0.7\arcsec$, respectively \citep[][]{KOEKEMOER07,AIHARA18}.

In this paper, we measure and analyze stellar proper motions in the COSMOS field. We do this by first bringing the ACS/F814W and HSC $i-band$ data which have an epoch differential of about 11 years, into astrometric concordance. We use {\it Gaia} Data Release 2 \citep[DR2]{GAIA-DR2} sources within COSMOS, as a benchmark comparison of our measured proper motions at bright magnitudes. The greater sensitivity of ACS and HSC photometry affords the potential to increase the number of proper motion sources in the field, and to extend the volume covered by them. 

Our aim for these studies is to understand the spatial distribution and velocities of a large sample of faint stars in the line of sight towards the COSMOS field. The proper motions, coupled with optical colors and information on extinction, allow us to estimate distances to and absolute magnitudes of these faint stars, using extrapolations from the known parallaxes of brighter Gaia DR2 sources. Therefore we can study the velocities of many faint stars as a function of spectral type and luminosity class. \citet{GRILLMAIR17} discovered ancillary Galactic streams within the Orphan Stream complex in the Sagittarius South arm, using Sloan Digital Sky Survey \citep[SDSS;][]{SDSS} data. In particular, the Sangarius stream within this complex partially crosses the COSMOS field. \citet{GRILLMAIR17} did not detect sources in {\it Gaia} Data Release 1, in the Sangarius stream, because of insufficient sensitivity. One of our goals is to search for members of this stream in both Gaia DR2 and in JSP by leveraging the more numerous and fainter sources which have astrometry measured through the latter. \citet{QIU2021} analyzed stellar proper motions at distances up to $\sim$100 kpc in an area of $\sim$100 square degrees, including part of the Sagittarius stream, by cross-correlating HSC and SDSS astrometry. \citet{QIU2021} find that within $\sim$1 kpc, red stars show proper motions consistent with Galactic rotation. Further out, up to $\sim$5 kpc, proper motions decrease with increasing distance. At halo distances $\sim$10--40 kpc, the Sagittarius stream affects proper motions. Beyond $\sim$40 kpc, halo blue stars exhibit isotropic velocities. Our goals in this paper are analogous to the above study.

Section \ref{sec:observations} describes the JSP prototype data set from ACS, HSC, and
{\it Gaia} DR2 in COSMOS. Section \ref{sec:matches} describes how we spatially matched stellar sources between these data sets. Section \ref{sec:astrometry_corr} presents the astrometrically corrected JSP data, using the grids of either {\it Gaia} DR2 stars or empirically selected compact galaxies as references, and resulting JSP proper motions of sources matched to {\it Gaia} DR2 stars. Section \ref{sec:FS_wrt_Gaia} presents the proper motions of all JSP sources, of which Gaia DR2 matches are a subset, in the astrometric grid of Gaia DR2. Section \ref{sec:uncertainties} discusses the uncertainties in our derived JSP proper motions. Section \ref{sec:stellar_distances_velocities} presents our derivation of stellar space velocities after correcting proper motions for the motion of the Sun relative to the Local Standard of Rest (LSR). Section \ref{sec:discussion} discusses our results, and Section \ref{sec:conclusions} contains our conclusions.

\section{Observational Data From Hubble and Subaru} \label{sec:observations}

The COSMOS field covers an area between approximately R.A. 149.12$\arcdeg$--151.12$\arcdeg$ and Decl.\ 1.2$\arcdeg$--3.2$\arcdeg$ \citep[][]{SCOVILLE07}. The Subaru/HSC First Data Release of this field consists of images in 133 12 arcmin $\times$ 12 arcmin so-called ``patches'' \citep[][]{AIHARA18}. We utilized the coadded ``calexp'' images and catalogs of HSC, where each calexp coadd was obtained from 104 individual exposures (of 300 sec each) typically taken within the same night. Each ``calexp'' coadd corresponds to a ``patch.'' The epochs of each coadd of HSC are within the year 2015. 

The Hubble/ACS observations were taken from HST Cycles 12, 13 only, and span the epoch ranges 2003 Oct 15--2004 May 21 and 2004 Oct 15--2005 May 21. For these, we considered the coadded exposures as described in \citet{KOEKEMOER07}. Each coadded exposure and its catalog correspond to an ACS ``patch'' where its location and dimensions have an identical counterpart in HSC. The region of the COSMOS field covered by HSC and ACS common patches is $\sim$1.64 square degrees.

We measured the equatorial coordinates and magnitudes in the HSC {\it i}-band calexp coadd and ACS F814W-band coadd with SExtractor \citep{SEXTRACTOR}. Magnitudes in this paper are in the AB system \citep{AB_MAG_SYSTEM} unless otherwise noted. The chosen bandpasses are the most sensitive to point sources in these two instruments. The output coordinates from SExtractor were unweighted center-of-light measurements, with both aperture photometry and isophotal photometry available. The SExtractor HSC positions were recentered via 2-dimensional Gaussian fits because we found, upon visual examination of HSC point sources, that the PSFs exhibited elliptical shapes not always centered at the center-of-light of the respective source. ACS point source images did not exhibit such ellipticity or asymmetry. This visual examination of HSC sources and ACS sources revealed that saturation ensues at magnitudes of $\sim$18.5. The reduced $\chi^2$ of 2-dimensional Gaussian fits increases rapidly with increasing source brightness starting at $\sim$18.5 magnitudes, thus confirming the onset of saturation.

The SExtractor astrometry and photometry of unsaturated sources in ACS and HSC were used in our subsequent analysis, and astrometric corrections were further applied, as described in Section \ref{sec:astrometry_corr}.

\section{Astrometric Matching Between ACS, HSC, and Gaia DR2 In the COSMOS Field} \label{sec:matches}

In order to assess the astrometry of HSC and ACS that we obtained with SExtractor, we need to cross-correlate it to an independent ``standard of truth.'' We regarded {\it Gaia} as one such standard and describe it here and in Section \ref{sec:gaia_wrt_gaia}, while in Section \ref{sec:gaia_wrt_empAGN} we considered an alternative, relative standard, consisting of a sample of empirically selected compact galaxies. The COSMOS field was searched in the Gaia DR2 catalog, for single and non-variable sources (using the Gaia DR2 flag conditions {\it duplicated\_source} $=$ 0 and {\it phot\_variable\_flag} $=$ NOT\_AVAILABLE). Only sources with Gaia parallax $>$ 0 were selected, meaning they had a valid solution for proper motion and parallax. \citet{GAIADR2PARALLAX} discuss the biases induced when removing Gaia parallaxes $<$ 0 in Gaia sample selections. The effect is to favor the selection of stellar sources at closer distances, in detriment to those at larger distances from us. \citet{GAIADR2PARALLAX} plotted the histogram of the Gaia DR2 parallaxes of 556,849 quasars from the AllWISE survey and showed that their mean was, as expected for these distant sources, close to 0, namely $\sim -10 \mu$as. When removing negative parallaxes, the mean was unrealistically large, namely $\sim$0.8 mas. Our sample of sources is much larger than Gaia stars, as will be discussed in Section \ref{sec:FS_wrt_Gaia}, but we repeated all of our calculations that involved Gaia stars, for comparison, by including all Gaia DR2 parallaxes, even those $<$ 0. The effects of using all Gaia DR2 parallaxes on our analysis are presented in Sections \ref{sec:astrometry_corr}, \ref{subsec:centroid_astrometry}, and \ref{sec:stellar_distances_velocities}.

\begin{figure*}[htbp]
\includegraphics[angle=0, width=6.0in]{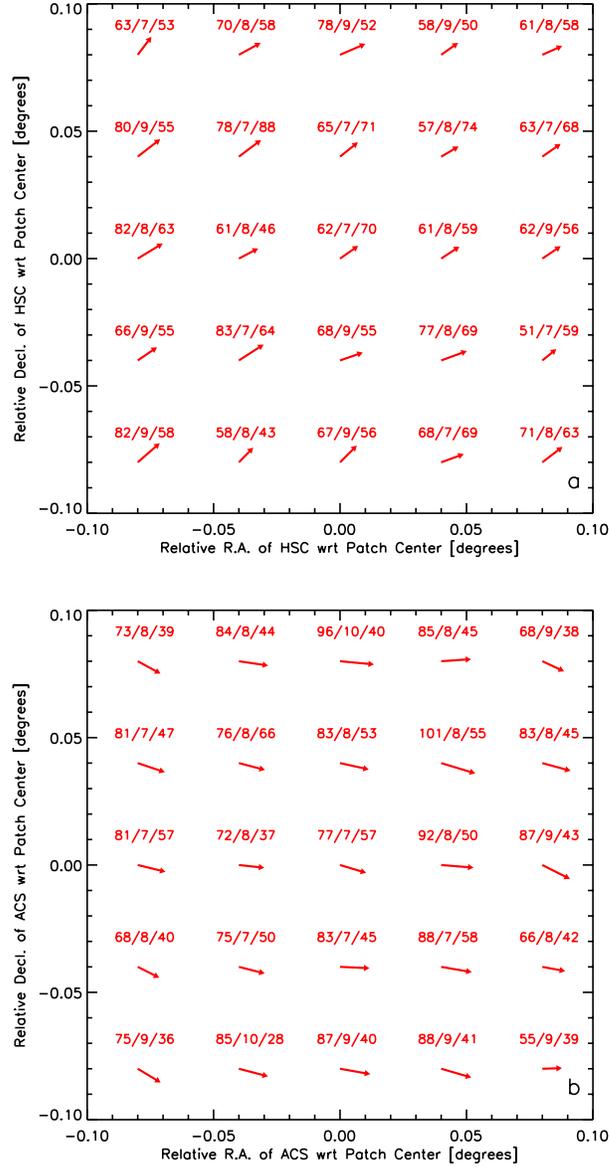}
\caption{\small (a) ({\it top panel}) The 63 HSC patches common to ACS were superposed or stacked, after transforming the equatorial coordinates of all HSC sources to relative coordinates with respect to the center of their respective patch. The {\it red arrows} are the median in each 2.4$\times$2.4 arcmin$^{2}$ cell within an HSC patch, of median offset vectors ${<}\Delta {\rm RA(HSC,Gaia)}{>}_{15}$ and ${<}\Delta {\rm Dec(HSC,Gaia)}{>}_{15}$ between HSC and Gaia DR2 matched pairs. ${<}\Delta {\rm RA(HSC,Gaia)}{>}_{15}$ and ${<}\Delta {\rm Dec(HSC,Gaia)}{>}_{15}$ were each computed from up to 15 matched pairs within 4 arcmin of the source, including the pair of the source itself and its Gaia DR2 match, as described in the text. Sets of three numbers above each vector list its magnitude (mas), 1-standard deviation of the mean (mas), and the number of median offset vectors in the cell. (b) ({\it bottom panel}) Similar to (a) above, the 63 ACS patches were superposed, and the {\it red arrows} are the median offset vectors in each cell within a patch, of median offset vectors ${<}\Delta {\rm RA(ACS,Gaia)}{>}_{15}$ and ${<}\Delta {\rm Dec(ACS,Gaia)}{>}_{15}$ between ACS and Gaia DR2 matched pairs. The sets of three numbers above each vector are also as described in (a) above.}
\label{fig:HSC_and_ACS_deltas_wrt_Gaia_5x5}
\end{figure*}

In the 133 HSC calexp coadds, 2,434 unique sources were spatially matched to Gaia DR2 stars, using a 0.2-arcsec radius circle. This search radius was chosen because it resulted in successful matches without source confusion and because the Gaia proper motions of unsaturated sources were $<<200$ mas/yr. Duplicate HSC sources in the areas of overlap of adjacent calexp coadds \citep[e.g.,][]{FAISST2021} were already removed. Among the above 133 HSC coadds, 63 were measured in common with ACS. In these 63 HSC coadds, 1,514 sources were matched to Gaia DR2 stars as described above. Similarly, 1,135 sources in the 63 ACS coadds were matched to Gaia DR2 stars. In each case the Gaia star positions were propagated to the epoch of each HSC patch, and to the mean epoch of the various individual exposures that covered each detected ACS source, using their Gaia-measured proper motions, before the above spatial match. Corrections for parallaxes obtained from the Gaia DR2 catalog were also applied during this process.

In Appendix \ref{sec:appendix_A} we describe the various samples of stars that we have used in our analysis throughout this paper.

For each HSC-Gaia matched pair, with separations in RA and Dec of respectively $\Delta{\rm RA(HSC,Gaia)}$, $\Delta{\rm Dec(HSC,Gaia)}$ (in mas) between the two components, the medians ${<}\Delta {\rm RA(HSC,Gaia)}{>}_{15}$ and ${<}\Delta {\rm Dec(HSC,Gaia)}{>}_{15}$ of this matched pair and up to 14 neighboring ones within 4 arcmin were computed. That is, the subscript ``15'' indicates that up to 15 offset vectors such as $\Delta{\rm RA(HSC,Gaia)}$ went into the computation of the median ${<}\Delta {\rm RA(HSC,Gaia)}{>}_{15}$. The HSC source was at the center of the 4-arcmin-radius circle containing its (HSC,Gaia) matched pair and up to 14 neighboring matched pairs. The number of neighbors and the size of the area searched for them were chosen so as to have a sufficient number of sources to calculate statistics and to have a small enough area so that the median offsets ${<}\Delta {\rm RA(HSC,Gaia)}{>}_{15}$ and ${<}\Delta {\rm Dec(HSC,Gaia)}{>}_{15}$ were representative of each particular sky position without cross-contamination from other areas. The smallest spatial scale of variations of median offset vectors is $\sim$2--3$\arcmin$, as found by visual inspection in both HSC and ACS coadds. The purpose of obtaining median offsets was to assess the astrometric difference between HSC and Gaia DR2 as a function of sky position, and to subsequently correct the HSC calexp coadd astrometry for any residuals not accounted for during the generation of these coadds. In order to illustrate the variations in median offsets and their dispersion in the COSMOS field, we superposed all 63 HSC patches (that were common to ACS) after transforming the coordinates within each patch to relative coordinates with respect to the patch center. The resulting ``stacked patch'' had much higher S/N in the median offset vectors than individual patches, and was divided into 5$\times$5 ``cells'' of 2.4$\times$2.4 arcmin$^{2}$ dimension each. Figure \ref{fig:HSC_and_ACS_deltas_wrt_Gaia_5x5}a shows the median, within a cell in the ``stacked'' HSC patch, of the median offset vectors ${<}\Delta {\rm RA(HSC,Gaia)}{>}_{15}$ and ${<}\Delta {\rm Dec(HSC,Gaia)}{>}_{15}$ between HSC and Gaia DR2. Similarly, Figure \ref{fig:HSC_and_ACS_deltas_wrt_Gaia_5x5}b shows the median, within a cell in the ``stacked'' ACS patch, of the median offset vectors ${<}\Delta {\rm RA(ACS,Gaia)}{>}_{15}$ and ${<}\Delta {\rm Dec(ACS,Gaia)}{>}_{15}$ between ACS and Gaia DR2. Figures \ref{fig:HSC_and_ACS_deltas_wrt_Gaia_5x5}a and \ref{fig:HSC_and_ACS_deltas_wrt_Gaia_5x5}b show that the median offset vector between HSC and Gaia DR2 has magnitude $\sim$65 mas and points to approximately the NE direction, while that between ACS and Gaia DR2 has magnitude $\sim$80 mas and points to approximately the ESE direction. The astrometric offset between each of the HSC and ACS data sets and Gaia DR2 is thus to first order a systematic translation. However, these figures also indicate variations from 2.4$\times$2.4 arcmin$^{2}$-cell to cell. The offset vector of HSC in the upper-left corner of Figure \ref{fig:HSC_and_ACS_deltas_wrt_Gaia_5x5}a is significantly ``steeper'' relative to the W-E direction than in most of the area of an HSC patch, while Figure \ref{fig:HSC_and_ACS_deltas_wrt_Gaia_5x5}b shows that the ACS offset vector in the lower-right corner of an ACS patch is practically ``horizontal'' (pointing to the East). These variations are due to residuals in the astrometric grids of HSC and ACS coadds, although they are not distortion corrections per se because HSC and ACS patches are not uniquely associated with detectors. Figures \ref{fig:HSC_and_ACS_deltas_wrt_Gaia_5x5}a and \ref{fig:HSC_and_ACS_deltas_wrt_Gaia_5x5}b illustrate the corrections that can be applied to the astrometric grids of HSC and ACS to bring them into alignment with the Gaia DR2 astrometric grid.

\section{Astrometric Corrections of HSC and ACS and Resulting JSP Proper Motions of Gaia DR2 Sources} \label{sec:astrometry_corr}

In this Section we correct the astrometry of HSC and ACS detected sources to the grids of either Gaia DR2 stars or empirically selected compact galaxies.

\subsection{Astrometric Correction to the Gaia DR2 Grid}\label{sec:gaia_wrt_gaia}

We applied astrometric corrections of ACS and HSC to the Gaia DR2 grid by subtracting, from SExtractor R.A. and Dec coordinates of each source, median offsets ${<}\Delta {\rm RA(HSC,Gaia)}{>}_{15}$, ${<}\Delta {\rm Dec(HSC,Gaia)}{>}_{15}$ (in the case of HSC), and ${<}\Delta {\rm RA(ACS,Gaia)}{>}_{15}$, ${<}\Delta {\rm Dec(ACS,Gaia)}{>}_{15}$ (in the case of ACS), described in Section \ref{sec:matches}.

\begin{figure*}[htbp]
\centering
\includegraphics[angle=0, width=4.7in]{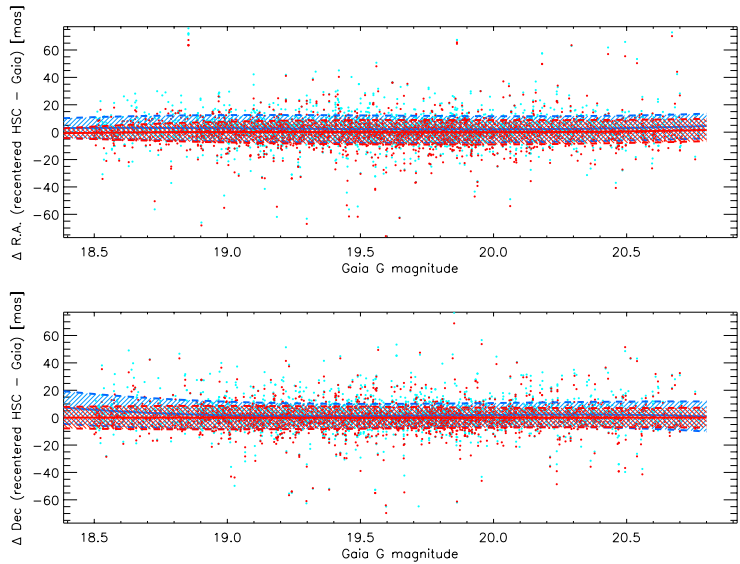}
\caption{Offsets in R.A., denoted as $\Delta {\rm RA(HSC,Gaia)}$ ({\it top panel}) or in Dec., denoted as $\Delta {\rm Dec(HSC,Gaia)}$ ({\it bottom panel}) between HSC SExtractor (recentered via 2-dimensional Gaussian fits) and Gaia DR2 catalog coordinate, as a function of $G$, for 2,434 sources. Astrometric corrections to the Gaia DR2 grid have been applied as explained in the text. The {\it small red symbols} are the offsets (in the sense HSC - Gaia) after propagating the positions of Gaia DR2 stars to the epochs of HSC coadds and applying Gaia DR2 catalog parallax corrections. The {\it solid red line} is the magnitude-binned 3$\sigma$-clipped mean of the offsets (using a 1-magnitude boxcar), smoothed with a 4th-order polynomial. The {\it red dashed lines} are the upper and lower 1-standard deviations of these offsets, also magnitude binned and smoothed as above, and the {\it red hashed region} represents the $\pm$ 1-standard deviation of these offsets. The {\it small blue symbols, lines, and hashed region} are similar offsets and their statistics, but without applying proper motion or parallax to Gaia stars.}
\label{fig:HSC_rdelt_vs_mag_Gaia}
\end{figure*}

Figure \ref{fig:HSC_rdelt_vs_mag_Gaia} shows offsets $\Delta {\rm RA(HSC,Gaia)}$ and $\Delta {\rm Dec(HSC,Gaia)}$ between HSC recentered (Section \ref{sec:observations}) and astrometrically corrected (to the Gaia DR2 grid) SExtractor sources and Gaia DR2 matched stars, as a function of Gaia $G$-magnitude (hereafter also referred to as $G$). Owing to the proximity of the epochs of Gaia and HSC observations, the application of proper motion has a modest but definite effect in reducing the standard deviations of the offsets from $\sim$15-20 mas in the case of no application of proper motion to $\sim$5-10 mas after applying proper motion. The offsets have a mean of zero after applying proper motion, as opposed to a mean of up to $\sim +$7 mas (at the brightest magnitudes) if not applying it. When comparing Figure \ref{fig:HSC_and_ACS_deltas_wrt_Gaia_5x5}a and Figure \ref{fig:HSC_rdelt_vs_mag_Gaia}, it can be seen that the application of astrometric correction to the Gaia DR2 grid leads to residual offsets of $\sim \pm$ 5 mas in either R.A. or Dec. vs. the original offsets of $\sim +$ 50 mas in either coordinate axis.

As an illustration of the effect of including all Gaia parallaxes, positive and negative, in the above astrometric corrections of HSC to the Gaia DR2 grid, we found a larger dispersion in the $\Delta {\rm RA(HSC,Gaia)}$ and $\Delta {\rm Dec(HSC,Gaia)}$ offsets at bright $G$ mags, of $\sim$10--15 mas as opposed to only $\sim$5 mas when only using positive parallaxes. The dispersion in offsets was comparable at $19<G<20$ for either positive or all Gaia parallaxes, at $\sim$5--10 mas. At $G$ of 20.7, the dispersion in Dec. was larger when using all Gaia parallaxes ($\sim$10 mas) than when using only positive ones ($\sim$5--6 mas).

\begin{figure*}[htbp]
\centering
\includegraphics[angle=0, width=5.7in]{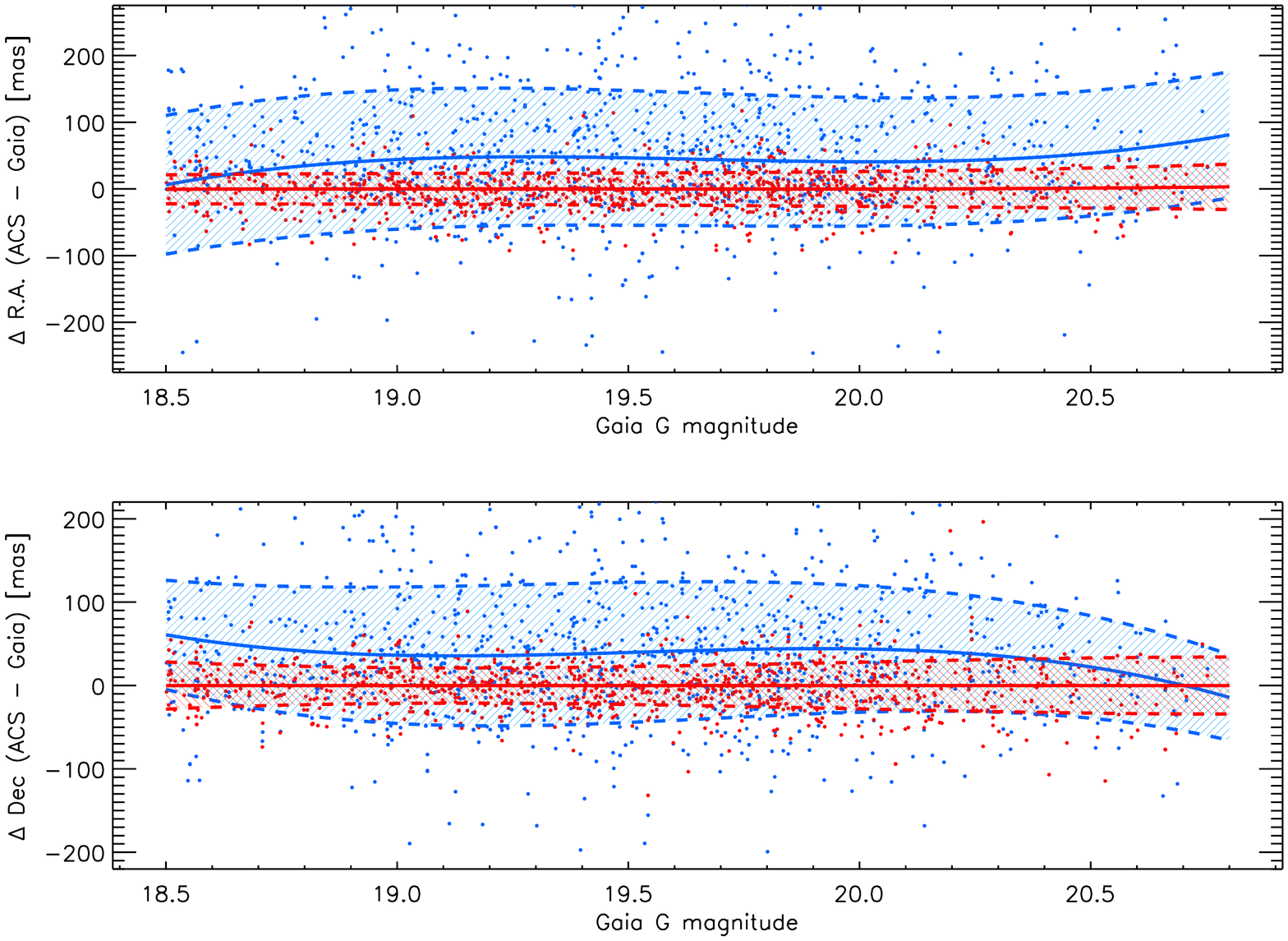}
\caption{Offsets in R.A., denoted as $\Delta {\rm RA(ACS,Gaia)}$ ({\it top panel}) or in Dec., denoted as $\Delta {\rm Dec(ACS,Gaia)}$ ({\it bottom panel}) between ACS SExtractor and Gaia DR2 catalog coordinate, as a function of $G$, for 1,135 sources. The {\it small red symbols} are the offsets (in the sense ACS - Gaia) after propagating the positions of Gaia DR2 stars to the mean epoch of all ACS single exposures that covered a given matched ACS source. The {\it solid red line} is the magnitude-binned 3$\sigma$-clipped mean of the offsets after applying proper motion (using a 1-magnitude boxcar), smoothed with a 4th-order polynomial. The {\it red dashed lines} are the upper and lower 1-standard deviations of these offsets, also magnitude binned and smoothed as above, and the {\it red hashed region} represents the $\pm$ 1-standard deviation of these offsets. The {\it small blue symbols, lines, and hashed region} are similar offsets and their statistics, but without propagating the positions of Gaia stars.} 
\label{fig:ACS_delta_vs_mag_Gaia}
\end{figure*}

Figure \ref{fig:ACS_delta_vs_mag_Gaia} shows offsets $\Delta {\rm RA(ACS,Gaia)}$ and $\Delta {\rm Dec(ACS,Gaia)}$ between ACS astrometrically corrected (to the Gaia DR2 grid) SExtractor sources and Gaia DR2 matched stars, as a function of $G$. In this case, owing to the $\sim$11-year difference in the epochs of ACS and Gaia matched stars, the application of proper motion to Gaia stars has a significant effect in reducing the standard deviation of the offsets from $\sim$100 mas in the case of no application of proper motion to $\sim$30 mas after applying proper motion. The offsets have a mean of zero after applying proper motion, just as in the case of HSC, as opposed to a mean of up to $\sim +$70 mas if not applying it. When comparing Figure \ref{fig:HSC_and_ACS_deltas_wrt_Gaia_5x5}b and Figure \ref{fig:ACS_delta_vs_mag_Gaia}, it can be seen that the application of astrometric correction to the Gaia DR2 grid leads to residual offsets of $\sim \pm$ 30 mas in either R.A. or Dec. vs. the original offsets of $\sim +$ 70--85 mas in R.A. and 0--$\sim -$37 mas in Dec.

Both the HSC-Gaia DR2 and ACS-Gaia DR2 residual offsets remain nearly constant with $G$. These residual offsets are most likely dominated by systematic effects of the astrometric grid of HSC and ACS, rather than by Poisson statistics of individual sources, as is further discussed in Section \ref{sec:uncertainties}.

In turn, the HSC-Gaia matched sources and the ACS-Gaia matched sources were cross-correlated, to thus identify 1,010 HSC-ACS matched sources (that is, HSC and ACS sources matched to a common Gaia DR2 star), of $G$ between 18.5 (that is, unsaturated; see Section \ref{sec:observations}) and 20.76. The proper motions of these sources were estimated using the SExtractor coordinates of HSC (recentered via 2-dimensional Gaussian fits) and ACS, both astrometrically corrected to the Gaia DR2 grid, and using the epochs of HSC observations listed in the calexp FITS headers, and the mean of the epochs of single-exposures of Cycles 12, 13 that covered each ACS source. These proper motions are referred to as ``JSP'' or ``HSC-ACS'' ones in what follows, to distinguish them from Gaia DR2 ones. JSP proper motions are not affected by the effect of differential chromatic refraction \citep{DCR}. HSC is equipped with an atmospheric dispersion corrector (ADC), which partially corrects for the effect, while ACS space-based data are not affected by the Earth's atmosphere.

\begin{figure*}[htb]
\centering
\includegraphics[angle=0, width=5.7in]{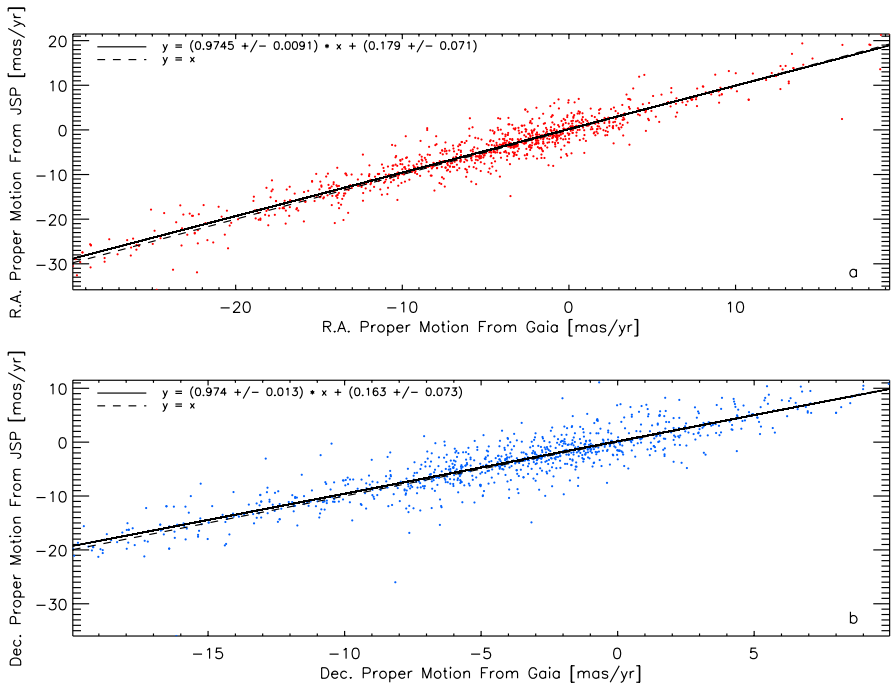}
\caption{Correlation of proper motions from JSP (estimated from HSC and ACS in this work) and from the Gaia DR2 catalog, for sources fainter than 18.5 Gaia DR2 magnitude. The {\it red symbols} are proper motions in R.A. and the {\it blue symbols} are proper motions in Dec. HSC and ACS astrometry was corrected to the grid of Gaia DR2 stars whose positions were propagated to the epochs of HSC and ACS, as explained in the text. The {\it solid lines} are least-squares linear fits, with parameters (slope, y-intercept, and standard deviations) listed in the Figure. The {\it dashed lines} have unity-slopes and represent perfect correlations. (a) The {\it top panel} shows the correlation of R.A. proper motion from JSP as a function of R.A. proper motion from Gaia DR2, for 972 sources with Gaia R.A. proper motion in the range $-$30 to $+$20 mas/yr. (b) The {\it bottom panel} shows the correlation of Dec. proper motion from JSP as a function of Dec. proper motion from Gaia DR2, for 925 sources with Gaia Dec. proper motion in the range $-$20 to $+$10 mas/yr.}
\label{fig:pm_jsp_and_gaia}
\end{figure*}

We compared the JSP proper motions of the 1,010 HSC-ACS-Gaia DR2 matches fainter than $G$ 18.5, with their Gaia DR2 catalog proper motions. Figures \ref{fig:pm_jsp_and_gaia}a and \ref{fig:pm_jsp_and_gaia}b show the correlation between R.A. and Dec. proper motions in JSP against Gaia DR2. A dashed line shows a 1-to-1 correlation, and a solid line shows a least-squares linear fit. It can be seen that JSP and Gaia DR2 proper motions are reasonably correlated; the fit has a slope within $\sim$3\% of a 1-to-1 correlation with uncertainty of $\sim$1\%, and the y-intercept is $\sim$0.16--0.18 mas/yr. In Section \ref{sec:uncertainties} we analyze what these statistics suggest for R.A. and Dec. JSP proper motions. We also explain in Section \ref{sec:uncertainties} why the majority but not all of the 1,010 sources were plotted in Figure \ref{fig:pm_jsp_and_gaia}.

The use of all Gaia parallaxes, regardless of whether they were positive or negative, yielded the same conclusions on the correlation of JSP and Gaia proper motions as above, when using only positive parallaxes.

\subsection{Astrometric Correction From Empirically Selected Compact Galaxies}
\label{sec:gaia_wrt_empAGN}

An alternative for astrometrically correcting JSP data is to use extragalactic point sources. Quasars, which would be optimal because they are stationary point sources, are very sparse, with only $\sim$1--3 expected in the COSMOS field. There are 297 unobscured broad-line (BLAGN) compact sources in the catalog of optical/IR counterparts \citep{MARCHESI16} to the 4,016 sources in the COSMOS-Chandra Legacy Survey \citep{CIVANO16}. Thus there are significantly fewer sources than Gaia stars. We generated a much larger sample by empirically selecting compact galaxies in the ACS coadds, where the source image ratio of semi-minor to semi-major axes is $>$ 0.9; SExtractor half-light radius is in the range 3--5 pixels (to exclude stars whose images are significantly smaller); SExtractor photometric S/N in the F814W filter is $>$ 25; and SExtractor magnitude in this bandpass is $>$ 20 (again to exclude bright stars and to avoid saturation). \citet{FAISST2021} describe the resulting ACS sample of 4,844 empirically selected compact galaxies.

\begin{figure*}[htbp]
\includegraphics[angle=0, width=6.0in]{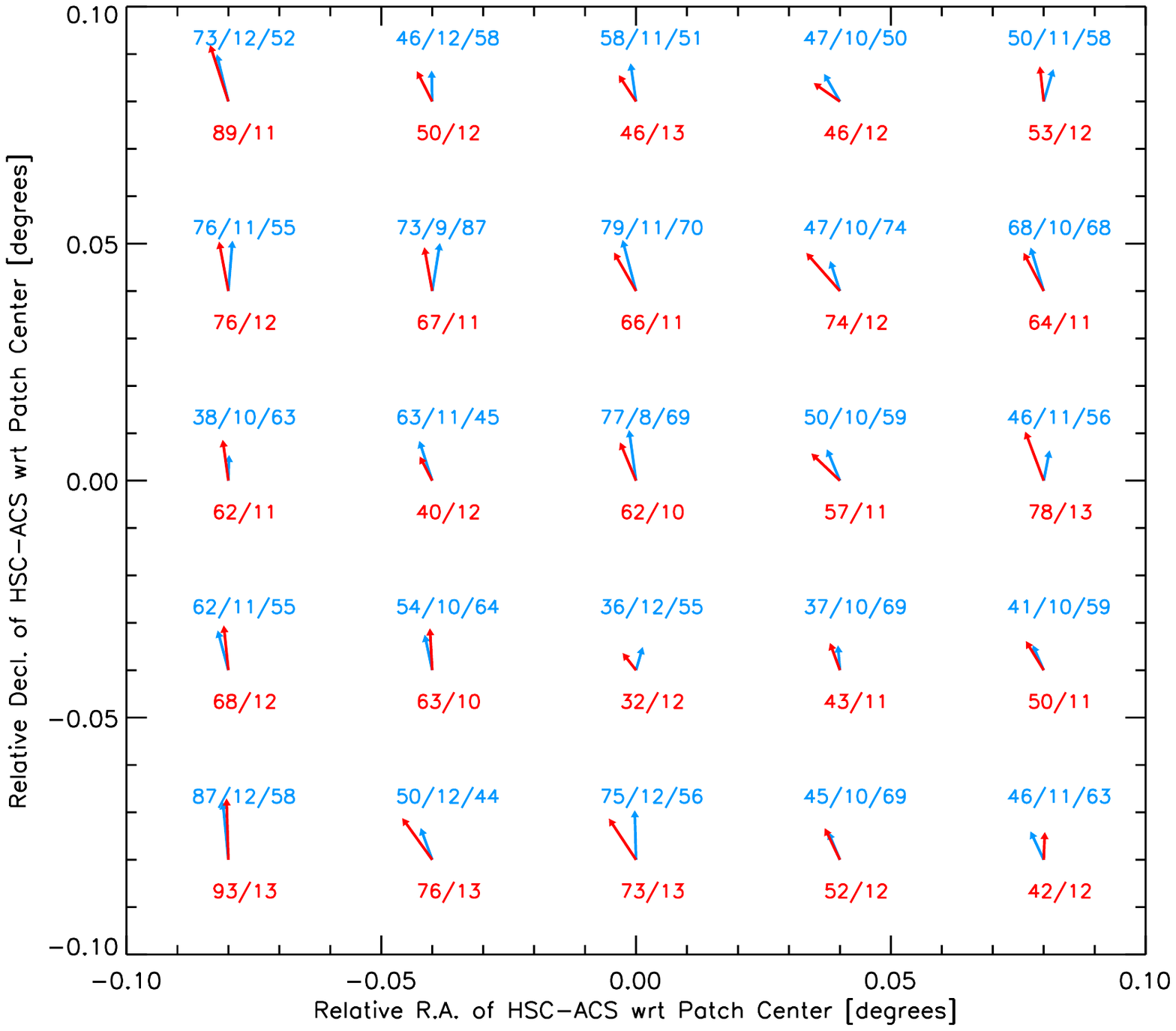}
\caption{The 63 HSC patches common to ACS were superposed or stacked, after transforming the equatorial coordinates of all HSC sources to relative coordinates with respect to the center of their respective patch (see Section \ref{sec:matches}). The {\it blue arrows} are the median in each 2.4$\times$2.4 arcmin$^{2}$ cell within the stacked HSC patch, of median offset vectors ${<}\Delta{\rm RA(HSC,ACS)}{>}_{15}$ and ${<}\Delta{\rm Dec(HSC,ACS)}{>}_{15}$ between HSC and ACS compact galaxy matched pairs. The median offset vectors were computed from up to 15 matched pairs within 4 arcmin of the source, including the pair of the HSC source itself and its ACS match, as described in the text. Sets of three numbers ({\it in blue font}) above each vector list its magnitude (mas), 1-standard deviation of the mean (mas), and the number of median offset vectors in the cell. For comparison, the {\it red arrows} are the median of the difference (separately in R.A. and in Dec.), within each cell, between median offset vectors ${<}\Delta{\rm RA(HSC,Gaia)}{>}_{15}$, ${<}\Delta{\rm Dec(HSC,Gaia)}{>}_{15}$ and median offset vectors ${<}\Delta{\rm RA(ACS,Gaia)}{>}_{15}$, ${<}\Delta{\rm Dec(ACS,Gaia)}{>}_{15}$. The sets of two numbers ({\it in red font}) above each vector list its magnitude (mas) and 1-standard deviation of the mean (mas).}
\label{fig:HSC_deltas_wrt_empAGN_5x5}
\end{figure*}

This sample was spatially matched to SExtractor HSC sources, using a 0.2-arcsec radius circle, to find 4,236 matches fainter than AB mag 18.5. As we do not expect a measurable mean motion for these extragalactic objects, the offsets between HSC and ACS should ideally be zero. In reality, owing to residuals not accounted for in the generation  of the coadds, non-zero offsets were found.

We denote by $\Delta{\rm RA(HSC,ACS)}$ and $\Delta{\rm Dec(HSC,ACS)}$ the separations in respectively R.A. and Dec. (in mas) between the two components of an HSC-ACS compact galaxy matched pair. For each matched pair, we then compute the median of $\Delta {\rm RA(HSC,ACS)}$ and those of up to 14 neighboring matched pairs within 4 arcmin and denote it by ${<}\Delta{\rm RA(HSC,ACS)}{>}_{15}$ (and a similar median in Dec. denoted by ${<}\Delta {\rm Dec(HSC,ACS)}{>}_{15}$). The purpose of obtaining these median offset vectors ${<}\Delta{\rm RA(HSC,ACS)}{>}_{15}$ and ${<}\Delta {\rm Dec(HSC,ACS)}{>}_{15}$ was, in analogy to Section \ref{sec:matches}, to assess the astrometric difference between HSC and ACS as a function of sky position and in particular to assess variations across the region covered by a 12$\times$12 arcmin$^{2}$ patch. These median offsets allow us to correct the HSC calexp coadd astrometry for any residuals not accounted for during the generation of the calexp coadds. That is, just as in Section \ref{sec:matches}, where Gaia DR2 stars were considered the ``standard of truth,'' in this case ACS compact galaxies were considered a relative standard (of HSC with respect to ACS). We superposed all 63 HSC patches common to ACS (Section \ref{sec:observations}), and divided the resulting ``stacked patch'' into 5$\times$5 ``cells.'' Figure \ref{fig:HSC_deltas_wrt_empAGN_5x5} shows the median within each cell of the median offset vectors ${<}\Delta{\rm RA(HSC,ACS)}{>}_{15}$ and ${<}\Delta {\rm Dec(HSC,ACS)}{>}_{15}$ for HSC-ACS compact galaxy matches ({\it blue arrows}). For comparison, Figure \ref{fig:HSC_deltas_wrt_empAGN_5x5} also shows, as ({\it red symbols}) the median within each cell of the difference (separately in R.A. and in Dec.) between median offset vectors ${<}\Delta{\rm RA(HSC,Gaia)}{>}_{15}$, ${<}\Delta{\rm Dec(HSC,Gaia)}{>}_{15}$, and median offset vectors ${<}\Delta{\rm RA(ACS,Gaia)}{>}_{15}$, ${<}\Delta{\rm Dec(ACS,Gaia)}{>}_{15}$. Figures \ref{fig:HSC_and_ACS_deltas_wrt_Gaia_5x5}a and \ref{fig:HSC_and_ACS_deltas_wrt_Gaia_5x5}b showed separately the medians within each cell of the two quantities that were subtracted, above. Ideally, if the astrometric corrections of HSC and ACS to the Gaia DR2 grid and the astrometric correction of HSC to the ACS compact galaxies grid were equivalent, then the relation $\Delta{\rm RA(HSC,Gaia)} - \Delta{\rm RA(ACS,Gaia)} = \Delta{\rm RA(HSC,ACS)}$ (and separately in Dec. also) would be exact and the {\it blue and red arrows} in Figure \ref{fig:HSC_deltas_wrt_empAGN_5x5} would be identical.

\begin{figure*}[htbp]
\centering
\includegraphics[angle=0, width=5.7in]{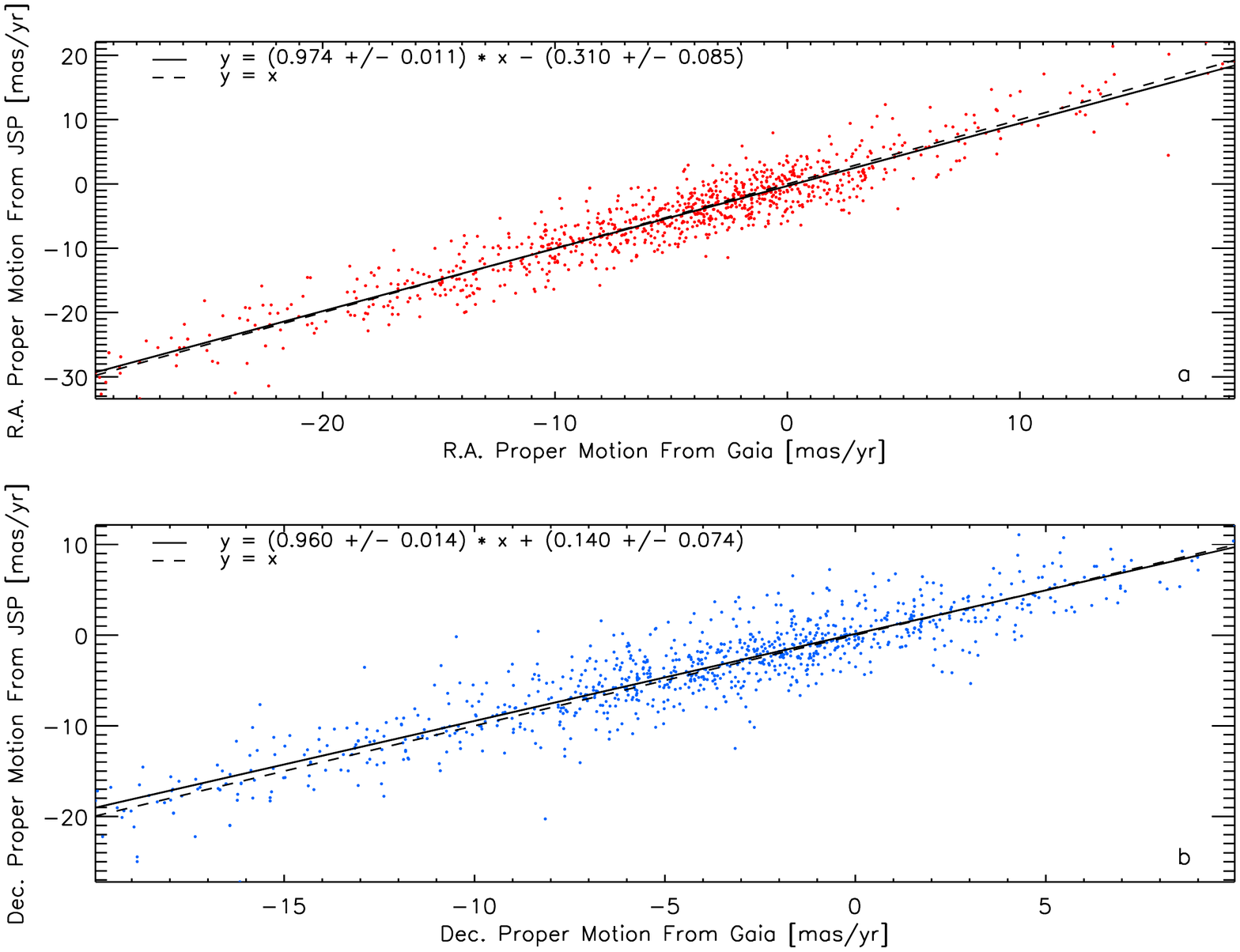}
\caption{Correlation of proper motions from JSP (estimated from HSC and ACS in this work) and from the Gaia DR2 catalog, for sources fainter than 18.5 Gaia DR2 magnitude. The {\it red symbols} are proper motions in R.A. and the {\it blue symbols} are proper motions in Dec. HSC astrometry was corrected to the grid of ACS compact galaxies described in the text and in \citet{FAISST2021}. The {\it solid lines} are least-squares linear fits, with parameters (slope, y-intercept, and standard deviations) listed in the Figure. The {\it dashed lines} have unity-slopes and represent perfect correlations. (a) The {\it top panel} shows the correlation of R.A. proper motion from JSP as a function of R.A. proper motion from Gaia DR2, for 971 sources with Gaia R.A. proper motion in the range $-$30 to $+$20 mas/yr. (b) The {\it bottom panel} shows the correlation of Dec. proper motion from JSP as a function of Dec. proper motion from Gaia DR2, for 925 sources with Gaia Dec. proper motion in the range $-$20 to $+$10 mas/yr.}
\label{fig:pm_jsp_vs_gaia_wrt_empAGN}
\end{figure*}

Figure \ref{fig:HSC_deltas_wrt_empAGN_5x5} shows that the {\it blue and red arrows} agree to within 10\% in magnitude but with excursions as high as 70\%, and within 15$\arcdeg$ in ``position angle'' but with excursions as high as 50$\arcdeg$. Therefore the astrometric corrections to the Gaia DR2 grid and to the grid of ACS compact galaxies are in reasonable agreement but the above differences are used in Section \ref{sec:uncertainties} to quantify the contribution of the astrometric grid correction uncertainty to that of JSP proper motions.

The median offset vector between HSC and ACS compact galaxies has magnitude $\sim$50 mas and points to approximately the NNW direction and is therefore to first order a systematic translation. However, there are variations across the region of patches. Four of the cells exhibit median per-source offset vectors pointing to the NNE direction. These variations are due to residuals in the astrometric grid of HSC relative to ACS, and Figure \ref{fig:HSC_deltas_wrt_empAGN_5x5} suggests corrections that can be applied to the HSC astrometric grid to bring it into relative alignment with the ACS astrometric grid. In this approach, the ACS grid would remain uncorrected.

We applied the astrometric correction of the HSC grid to the ACS grid of compact galaxies suggested by Figure \ref{fig:HSC_deltas_wrt_empAGN_5x5}, by subtracting the median offsets ${<}\Delta{\rm RA(HSC,ACS)}{>}_{15}$ and ${<}\Delta{\rm Dec(HSC,ACS)}{>}_{15}$ from the SExtractor coordinates of each HSC source. JSP proper motions of 1,010 Gaia DR2 stars were obtained as in Section \ref{sec:gaia_wrt_gaia}. We then cross-correlated these JSP proper motions to Gaia DR2 catalog proper motions. Figures \ref{fig:pm_jsp_vs_gaia_wrt_empAGN}a and \ref{fig:pm_jsp_vs_gaia_wrt_empAGN}b show these correlations in R.A. and Dec., respectively. We can see that the correlations are very comparable to those using astrometry corrected to the grid of Gaia DR2 stars (Figures \ref{fig:pm_jsp_and_gaia}a and \ref{fig:pm_jsp_and_gaia}b). In particular, the slopes of the correlations to Gaia DR2 catalog proper motions are within 1\% of each other in the two astrometric grids. The errors of the slope and y-intercepts agree to within 10\%. The y-intercepts are within $\pm$0.3 mas/yr although smaller in absolute value for the Gaia DR2 astrometric grid ($\sim$0.16--0.18 mas/yr) than the ACS compact galaxies grid ($\sim -0.31$--$+$0.14 mas/yr).

\section{Proper Motions of All JSP Sources, Astrometrically Corrected to the Gaia DR2 Grid}
\label{sec:FS_wrt_Gaia}

\begin{figure}[htbp]
\includegraphics[angle=0, width=3.2in]{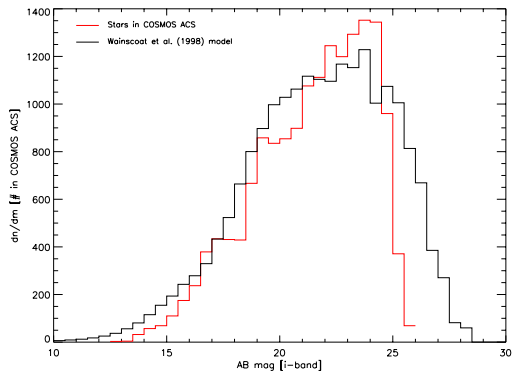}
\caption{The {\it black line} is the histogram of expected stellar source count as a function of AB mag in the COSMOS field, computed from the model by Wainscoat et al. (1998). The {\it red line} is the number of empirically determined stars in this field detected by ACS, as described in the text, and a separate sample detected by ACS and matched to Gaia DR2.}
\label{fig:predicted_vs_actual_starcounts}
\end{figure}

The astrometric potential of JSP is realized by measuring proper motions of a vast number of stellar sources fainter than Gaia. To predict the number of expected sources, we ran the Wainscoat et al. (1998) star counts model for the COSMOS field (Figure \ref{fig:predicted_vs_actual_starcounts}, {\it black line}). This model has been scaled to the common area covered by HSC and ACS (Section \ref{sec:observations}). We find that the total number of stars expected from the model, brighter than 25 AB mag is 1.7$\times10^{4}$ and that between 19 and 25 AB mag is 1.3$\times10^{4}$.

\begin{figure}[htbp]
\includegraphics[angle=0, width=3.2in]{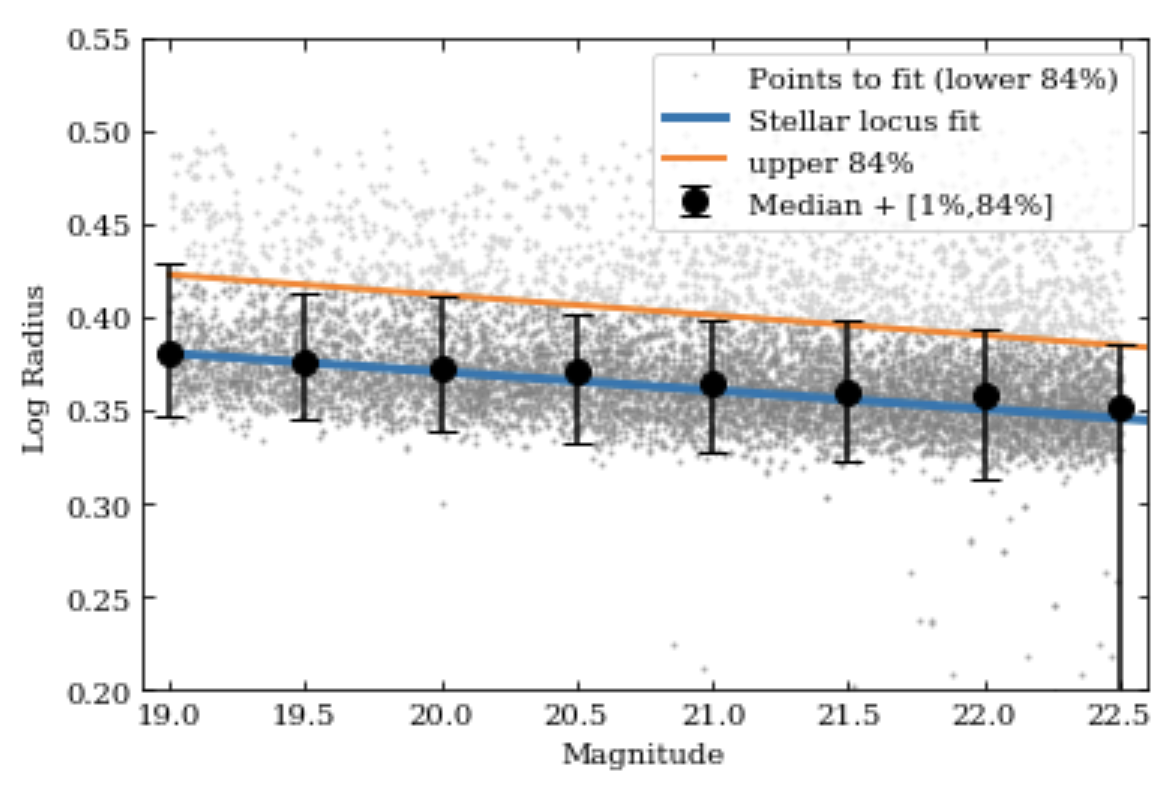}
\caption{The plot shows the technique used for empirical selection of faint stars. The y-axis shows the SExtractor half-light source radius (in pixels). A linear fit ({\it blue line} was obtained to the magnitude-binned median half-light source radius (in pixels) as a function of SExtractor F814W ACS AB magnitude. The error bars are the magnitude-binned 1- and 84-percentiles of the distribution. The {\it orange line} is a linear fit to the 84-percentiles, and sources below it are considered to be stars. }
\label{fig:empFS_radius_threshold}
\end{figure}

In order to build a large sample of stars in the COSMOS field, we first identified 13,009 sources in the ACS coadds that were empirically determined to be stars (Figure \ref{fig:predicted_vs_actual_starcounts}, {\it red line}). These sources have SExtractor magnitudes in the F814W filter between 19 and 26; they have a ``round'' morphology (i.e., the ratio of semi-minor to semi-major axes is $>$ 0.9); their SExtractor flag CLASS\_STAR is $>$ 0.9; and they satisfy SExtractor S/N $>$ 30. In addition, these sources were required to be located, in a plot of SExtractor half-light radius of the source image in the ACS coadds, as a function of SExtractor magnitude (Figure \ref{fig:empFS_radius_threshold}), below the 84-percentile of the distribution that was fit in the range $19-22.5\,{\rm AB}$. 

We added, to the sample of ACS empirically selected stars, 3,937 ACS stars matched to Gaia DR2 stars, unconstrained in apparent magnitude or parallax. From this sample, we then removed 453 sources that were already included in the empirically selected stars. Otherwise, the remaining ACS stars matched to Gaia DR2 stars were not in the ACS empirically selected stars mainly because they are brighter than 19 AB magnitude. Thus our starting large sample of ACS stars consists of 16,493 sources (Figure \ref{fig:predicted_vs_actual_starcounts}, {\it red line}). The number of these stars brighter than 25 AB mag is 1.6$\times10^{4}$ and that between 19 and 25 AB mag is 1.3$\times10^{4}$. These numbers are in very good agreement (10\% or better) with the above predictions from the Wainscoat model. The histogram of our large sample of ACS stars is shown as a {\it red line} in Figure \ref{fig:predicted_vs_actual_starcounts}, where the agreement with the Wainscoat model is 10\% or better up to AB mag 25. Stars fainter than 25 in our sample precipitously drop in numbers because of our S/N SExtractor threshold.

In the HSC SExtractor catalog, we spatially matched 11,783 sources to the 13,009 ACS empirically selected faint stars, using a 0.2-arcsec radius circle. The HSC and ACS positions were previously astrometrically corrected to the Gaia DR2 reference, as explained for Gaia-matches in Section \ref{sec:gaia_wrt_gaia}. We then computed their JSP proper motions and added them to the sample of 1,010 JSP proper motions of Gaia stars with positive parallaxes (Section \ref{sec:gaia_wrt_gaia}). Excluding 264 JSP proper motions of Gaia stars common to the above samples, we finally have a sample of 12,529 sources fainter than AB 18.5 mag with well-determined JSP proper motions.

Of the 1,010 Gaia-ACS stars with well-determined proper motions in the above sample, 746 were not detected in our empirical selection of faint stars. The reason is that $\sim$80\% of the 1,010 Gaia-ACS sources had SExtractor AB mag brighter than 19, thus being excluded by definition from the empirically selected faint stars. Of the remaining $\sim$20\%, half of them had anomalously high noise values, and the remaining half did not satisfy the ``roundness'' criterion described above.

In Sections \ref{sec:uncertainties} and \ref{sec:stellar_distances_velocities} we analyze the uncertainties of our derived JSP proper motions for the above large sample of stars, and our derivation of velocities and distances to these stars, respectively.

An alternative to the astrometric correction to the Gaia DR2 grid is to correct to the grid of 4,844 empirically selected compact galaxies, as described in \citet{FAISST2021}; see Section \ref{sec:gaia_wrt_empAGN}. Section \ref{sec:gaia_wrt_empAGN} showed that this correction, when applied to Gaia stars, yields comparable results to a correction to the Gaia DR2 grid. We also applied such correction to our set of 11,519 empirically selected faint stars and derived JSP proper motions very similar to those when corrected to the grid of Gaia DR2.

Owing to the similarity of JSP proper motions using either the grid of Gaia DR2 stars or that of compact galaxies, we confine our remaining analysis to the former. However, we still utilize the latter (correction to the grid of ACS compact galaxies) to estimate systematic uncertainties of the astrometric grid correction, as explained next in Section \ref{sec:uncertainties}.

\section{Uncertainties In JSP Proper Motions}
\label{sec:uncertainties}

In this Section we analyze the uncertainties in JSP proper motions using several approaches: (i) From the comparison of JSP and Gaia DR2 proper motions described in Section \ref{sec:gaia_wrt_gaia}); (ii) by combining Monte Carlo simulations of centroiding uncertainties and differences in astrometric corrections to the grids of Gaia DR2 and of empirically selected compact galaxies (Sections \ref{sec:gaia_wrt_gaia} and \ref{sec:gaia_wrt_empAGN}); (iii) by computing the residuals of JSP ``null motions'' of stationary extragalactic sources.

\subsection{Monte Carlo Simulations of JSP vs. Gaia Proper Motions}\label{subsec:montecarlo}

In Section \ref{sec:gaia_wrt_gaia} we compared JSP and Gaia DR2 catalog proper motions and the results were shown in Figure \ref{fig:pm_jsp_and_gaia}. We carried out Monte Carlo simulations of Figures \ref{fig:pm_jsp_and_gaia}a and \ref{fig:pm_jsp_and_gaia}b, separately, by generating 50 Gaussian-distributed random proper motion deviants relative to a perfect correlation of JSP and Gaia DR2 proper motions. Our free parameter was the standard deviation of JSP proper motion distributions. The standard deviations of Gaia proper motions were those from the Gaia DR2 catalog. For each JSP standard deviation, least-squares linear fits were estimated to the resulting Monte Carlo simulated JSP vs. Gaia proper motion correlation. Our goal was to match the simulation linear fit parameters to those in Figures \ref{fig:pm_jsp_and_gaia}a and \ref{fig:pm_jsp_and_gaia}b. Among the various simulations we tried, standard deviations in R.A. and Dec. JSP proper motion of 2.0 mas/yr each yield the closest match to the actual proper motion linear fits. These uncertainties are per axis (either R.A. or Dec.) and apply to all sources; therefore they are an approximation.

\subsection{Combination of Centroiding and Astrometric Grid Correction Uncertainties}\label{subsec:centroid_astrometry}

An estimate of uncertainty in proper motion by combining the uncertainty of centroiding and that from systematics in the correction to the grid of Gaia DR2 may be more accurate than that from the linear correlation of JSP and Gaia described above.

\begin{figure}[htbp]
\includegraphics[angle=0, width=3.2in]{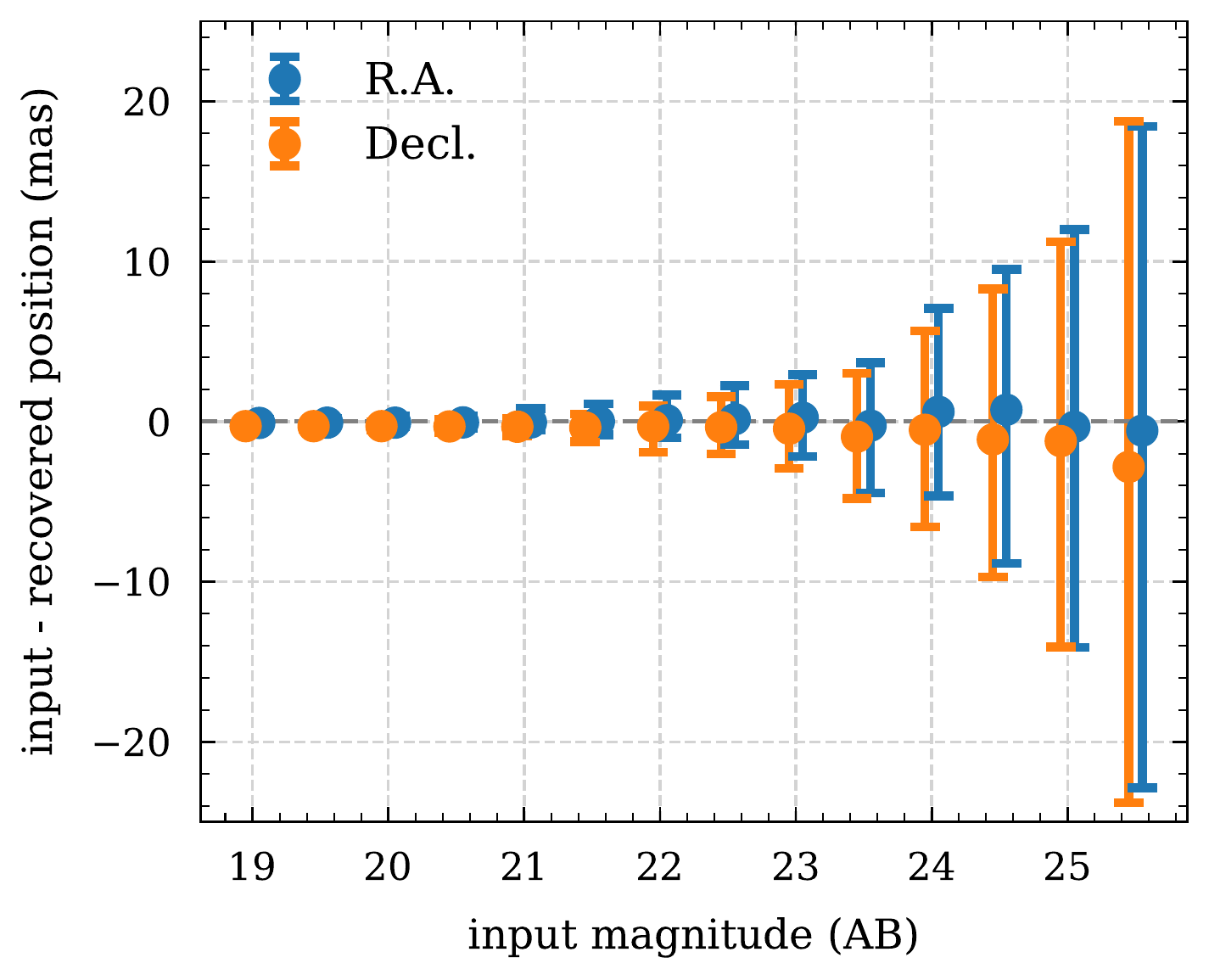}
\caption{Monte Carlo simulation of centroiding uncertainty in HSC. Plotted are the magnitude-binned positional differences between randomly generated point sources and their SExtractor recovered positions, separately in R.A. ({\it orange symbols}) and Dec. ({\it blue symbols}), as a function of SExtractor HSC-i magnitude. The error bars are 1-standard deviations of these differences, in each magnitude bin.}
\label{fig:astrometry_sim}
\end{figure}

The centroiding uncertainty of HSC was estimated by running Monte Carlo simulations on one patch, namely ``(5,4)'' (see \citet[][]{AIHARA18} for an explanation of patch nomenclature). Detected objects from SExtractor were masked and 10,000 random positions were generated, falling outside the masked positions. PSFs in the magnitude range 19--26.5 were generated and placed at the above random positions, and SExtractor was run on these artificial sources. The difference between the input random position and the SExtractor position for each artificial source was computed. The magnitude-binned standard deviations of these differences are shown as error bars in Figure \ref{fig:astrometry_sim}. We regard these uncertainties as representing the centroiding uncertainty of HSC. The corresponding centroiding uncertainty of ACS is $\sim$1/6 that of HSC, because this is the ratio of pixel sizes in HSC and ACS (0.168 arcsec and 0.030 arcsec, respectively).

The contribution of centroiding uncertainty to JSP proper motion uncertainty is estimated from Figure \ref{fig:astrometry_sim}. Since there are $\sim$11 years separating the epochs of ACS and HSC then, for a given HSC centroiding uncertainty $\sigma_{HSC,centr}$ as in the error bars in Figure \ref{fig:astrometry_sim}, the corresponding uncertainty in JSP proper motion, which includes ACS centroiding uncertainty, is 

\begin{equation}
\sigma_{JSP,centr} = \frac{\sigma_{HSC,centr}}{11}\sqrt{1 + (\frac{1}{6})^{2}} =  \frac{\sqrt{37}}{66}\sigma_{HSC,centr}
\label{eq:sigma_pm_centr}
\end{equation}

\begin{figure*}[htbp]
\centering
\includegraphics[angle=0, width=5.7in]{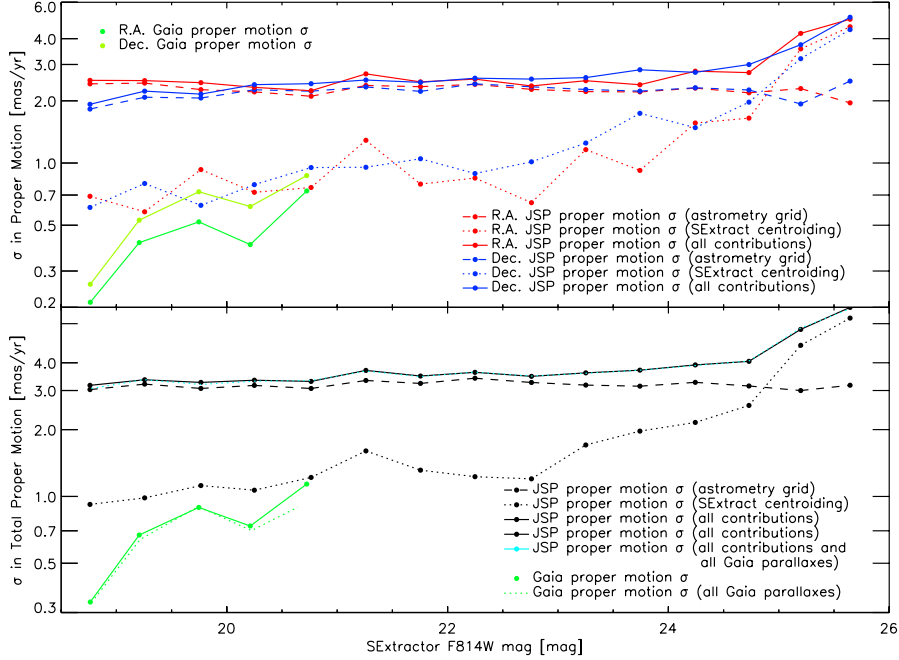}
\caption{Uncertainties in JSP proper motions. The {\it upper panel} shows uncertainties in each axis ({\it red symbols connected with red lines} are for R.A., and {\it blue symbols connected with blue lines} are for Dec.). For comparison, Gaia DR2 catalog proper motion uncertainties are shown as {\it aquamarine symbols connected with aquamarine lines} for R.A. and {\it green symbols connected with green lines} for Dec. These uncertainties were computed in 0.5-mag wide bins, where the photometry is from SExtractor in the ACS F814W bandpass. The JSP uncertainties from centroiding are shown as {\it symbols connected with dotted lines}, and the uncertainties from astrometric grid correction of HSC and ACS patches, as explained in the text, are shown as {\it symbols connected with dashed lines}. The JSP uncertainties from both of these contributions are shown as {\it symbols connected with solid lines}. The {\it lower panel} shows uncertainties in total (R.A. and Dec. combined in quadrature) JSP ({\it black symbols}) and Gaia DR2 catalog ({\it green symbols}) proper motions. The {\it lower panel} also shows, for comparison, uncertainties in total JSP ({\it light blue symbols connected with a dotted line}) and Gaia DR2 catalog ({\it green dotted line}) proper motions, when using all Gaia stars regardless of whether their parallaxes were positive or negative.} 
\label{fig:proper_motion_sigmas}
\end{figure*}

These uncertainties in proper motion arising from centroiding are shown in Figure \ref{fig:proper_motion_sigmas} as symbols connected with dotted lines. They generally increase with increasing magnitude, from $\sim$0.9 mas/yr to $\sim$6.2 mas/yr in total JSP proper motion.

We estimated the uncertainty of JSP proper motions arising out of systematics in astrometric grid correction, by finding the JSP proper motion difference for each source after correcting to the Gaia DR2 grid (Sections \ref{sec:gaia_wrt_gaia} and \ref{sec:FS_wrt_Gaia}) and after correcting to the grid of empirical compact galaxies (Section \ref{sec:gaia_wrt_empAGN}). These differences were binned by SExtractor ACS magnitude (in the F814W bandpass), and their standard deviations as a function of magnitude are shown in Figure \ref{fig:proper_motion_sigmas}, as symbols connected by {\it dashed lines}. These standard deviations represent the portion of uncertainty of JSP proper motions arising out of the uncertainty of astrometric correction of HSC and ACS (i.e., when comparing the astrometric corrections with respect to Gaia DR2, Section \ref{sec:FS_wrt_Gaia}, and with respect to empirical compact galaxies, Section \ref{sec:gaia_wrt_empAGN}). Possible sources of these systematic uncertainties are uncertainties in the Gaia DR2 proper motions, and contamination of the sample of compact galaxies with some stellar objects. The systematic uncertainties in astrometric grid correction are $\sim$2.0-3.05 mas/yr in R.A. and nearly constant with magnitude or $\sim$1.9--2.1 mas/yr in Dec. The larger variation in R.A. proper motion uncertainty, increasing towards brighter sources, is due to the systematic uncertainty in 2-dimensional Gaussian fitting of sources close to saturation, and due to the fact that the absolute value of astrometric corrections in R.A. relative to Gaia DR2 are larger than in Dec. (Figures \ref{fig:HSC_and_ACS_deltas_wrt_Gaia_5x5}a and b). The total contribution to JSP proper motion uncertainty (quadrature combination of R.A. and Dec. components) from astrometric grid correction is in the range $\sim$3.0--3.9 mas/yr.

We believe that the total uncertainties in JSP proper motions, from both centroiding and astrometric grid corrections, and applying to the quadrature combination of R.A. and Dec. proper motions, are as shown by symbols connected with solid lines in the {\it bottom panel} of Figure \ref{fig:proper_motion_sigmas}. They generally increase with increasing magnitude and range from $\sim$3 to $\sim$4 mas/yr in the magnitude range 18.5--25. The portion of uncertainty from centroiding starts to dominate over astrometric grid correction at $\sim$25 mag and the total uncertainty is as high as $\sim$6.5 mas/yr at 26 mag. We therefore confined our analysis to stars as faint as 25 mag.

The {\it bottom panel} of Figure \ref{fig:proper_motion_sigmas} also shows, for comparison, the total uncertainties in JSP proper motions, shown as {\it light blue symbols connected with a light blue dotted line}, when including all Gaia DR2 stars, irrespective of whether their parallaxes were positive or negative. The total uncertainties are almost indistinguishable when including all Gaia parallaxes during the astrometric grid correction, relative to when including only positive parallaxes. The {\it bottom panel} of Figure \ref{fig:proper_motion_sigmas} also shows the Gaia DR2 catalog proper motion uncertainties, as a {\it green dotted line}, for all Gaia parallaxes including negative ones. These uncertainties are smaller than when including only positive Gaia parallaxes, particularly for the faintest Gaia stars.

\subsection{Measuring Null Motions of Extragalactic Point Sources}\label{subsec:pseudo_pm}

\begin{figure*}[htbp]
\centering
\includegraphics[angle=0, width=5.7in]{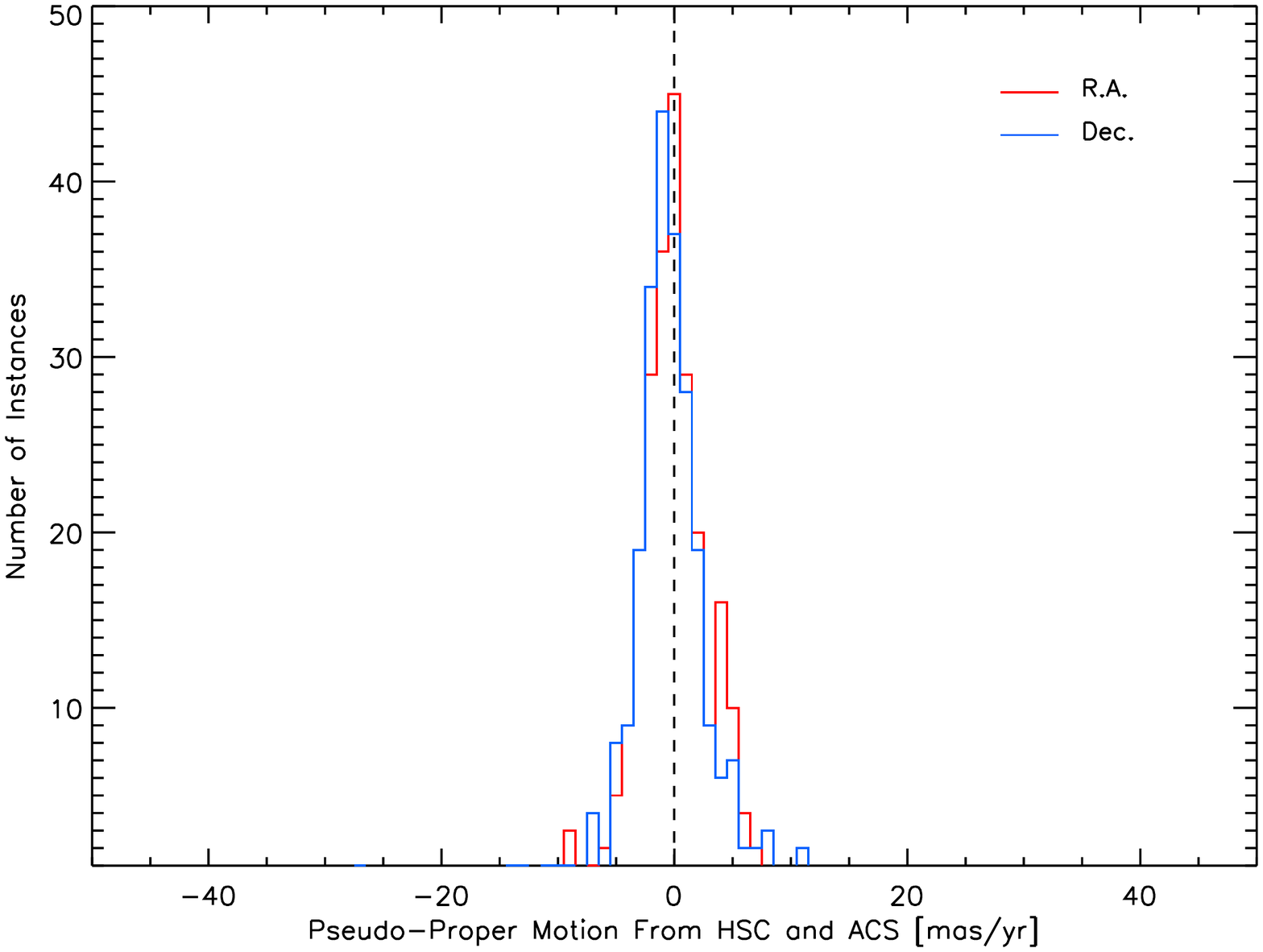}
\includegraphics[angle=0, width=5.7in]{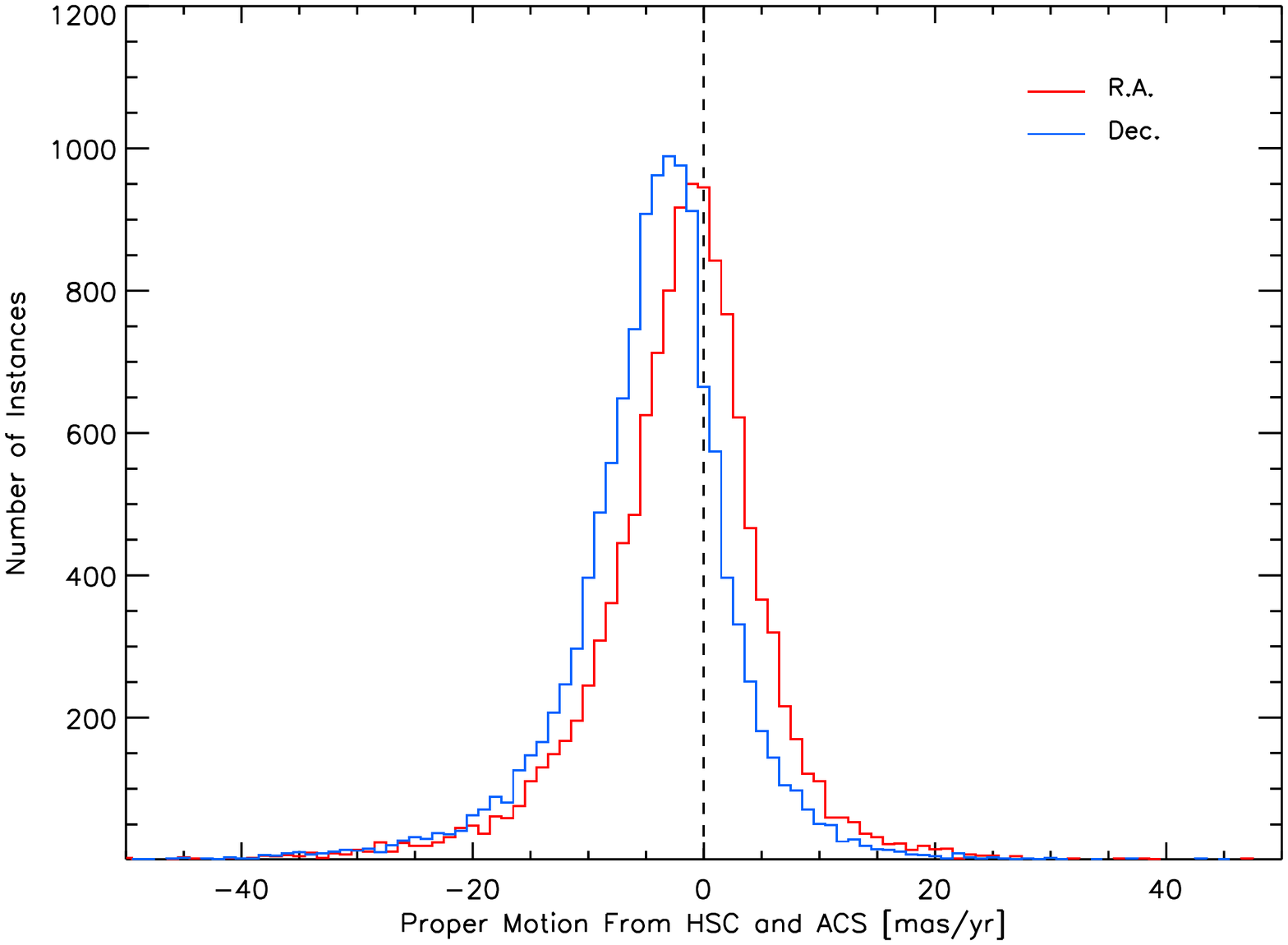}
\caption{{\it (a) Top panel}: histogram of JSP pseudo-proper motions of 296 BLAGN, explained in the text. The {\it red line} is for R.A. pseudo-proper motions and the {\it blue line} is for Dec. ones, both of which were expected to be statistically null. The {\it vertical dashed line} denotes zero proper motions. {\it (b) Bottom panel}: histogram of the JSP proper motions of 12,529 stars, drawn from Gaia DR2 (Section \ref{sec:gaia_wrt_gaia}) or from empirically selected faint stars (Section \ref{sec:FS_wrt_Gaia}). The {\it red, blue, and dashed lines} are as in {\it (a)} above. }
\label{fig:fake_BLAGN_and_true_empFS_proper_motions}
\end{figure*}

A third method to estimate the uncertainties of JSP proper motions is to astrometrically correct the HSC and ACS positions of unobscured broad-line AGN (BLAGN), and compute ``pseudo-proper motions'' by subtracting the HSC and ACS positions and dividing by their epoch difference, just as we did for stellar sources. The sample of 296 BLAGN \citep{CIVANO16,MARCHESI16} is as described in the beginning of Section \ref{sec:gaia_wrt_empAGN}. Figure \ref{fig:fake_BLAGN_and_true_empFS_proper_motions}a ({\it top panel}) shows a histogram of the JSP pseudo-proper motions of BLAGN. The R.A. pseudo-proper motion distribution peaks at 0 mas/yr while the Dec. one peaks at $\sim$1-2 mas/yr. Both distributions have FWHM $\sim$3--4 mas/yr, comparable to our other estimates of uncertainties. 

For comparison, Figure \ref{fig:fake_BLAGN_and_true_empFS_proper_motions}b ({\it bottom panel}) shows a histogram of the JSP proper motions of 12,529 stars (both Gaia DR2 sources, Section \ref{sec:gaia_wrt_gaia} and empirically selected faint stars, Section \ref{sec:FS_wrt_Gaia}). The R.A. proper motion distribution peaks at $\sim -$1 mas/yr with a FWHM of $\sim$10 mas/yr, while the Dec. one peaks at $\sim -$4 mas/yr with a FWHM of $\sim$12 mas/yr. Both the R.A. and Dec. distributions are asymmetrical and their FWHM are a factor of $\gtrsim$ 3$\sigma$ of the ``null'' pseudo-proper motions of BLAGN.

We conclude that the recovery of ``null'' proper motions of BLAGN (whose distribution peaks at zero within uncertainties), and the fact that these  uncertainties are comparable to those from our other methods (Sections \ref{subsec:montecarlo} and \ref{subsec:centroid_astrometry} shows the robustness of our method of JSP proper motion estimation.

\section{Obtaining Stellar Space Velocities in the Galaxy from JSP Proper Motions and Empirically Derived Distances} \label{sec:stellar_distances_velocities}

In order to derive space velocities of stars in our sample, using also estimates of their distances to the Earth, their proper motions need to be corrected for the effects caused by the motion of the Solar System relative to the LSR. This motion is in the apparent ``apex'' direction R.A.$_{\rm apex} = 277.0\arcdeg$, 
Dec.$_{\rm apex} = 30.0\arcdeg$, at a speed $\rm S_{\Sun} \sim$ 16 km/s. The effect of the solar motion on measured proper motions is to impart a component of motion that is maximal at sources located in a direction perpendicular to the above ``apex.'' This spurious component is in a direction away from the ``apex,'' along a ``meridian'' in a fictitious celestial coordinate system where the ``north pole'' is at the ``apex.'' \citet{MIHALAS_BINNEY} explain that the spurious component of motion has a speed of $\pi \rm S_{\Sun} \sin(\gamma) / 4.74$, where $\gamma$ is the great-circle angular distance between the ``north pole'' or ``apex'' and the source, and $\pi$ is its parallax in arcsec.

\begin{figure}[htbp]
\centering
\includegraphics[angle=0, width=3.2in]{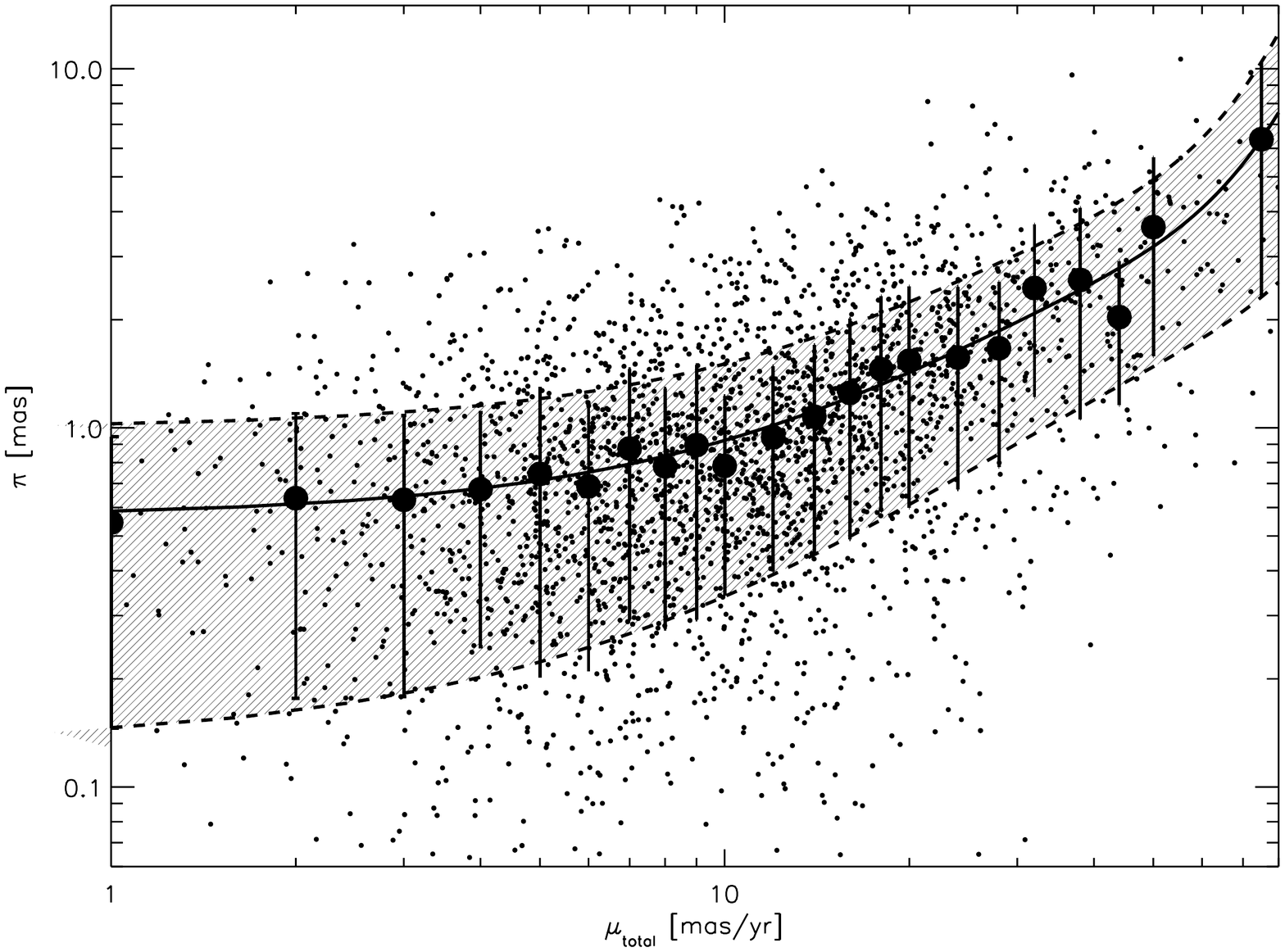}
\caption{Gaia DR2 catalog parallax ($\pi$) as a function of total proper motion ($\mu_{total}$) for 3,367 stars matched to ACS, and with parallaxes $>$ 0 indicating valid Gaia DR2 solutions. The {\it small black symbols} correspond to individual stars. The {\it large symbols with error bars} are the proper-motion-binned means and standard deviations of parallax. The {\it solid line} is a smoothed 4th order polynomial, described in the text (equation \ref{eq:prov_pi_vs_mu}, fitted to the average parallax values, and the {\it dashed lines} are similar fits to the $\pm$ 1-standard deviation values, which bound the {\it shaded region}.}
\label{fig:parallax_vs_motion}
\end{figure}

The parallax of sources was obtained from the Gaia DR2 catalog, or from an empirical relation between parallax and total proper motion derived from Gaia DR2 matched sources. Figure \ref{fig:parallax_vs_motion} shows the Gaia DR2 parallax of 3,367 sources with positive values
(indicating valid Gaia DR2 solutions), out of the 3,937 Gaia matches to ACS sources, as a function of total Gaia DR2 proper motion (quadrature combination of the R.A. and Dec. components).

From Figure \ref{fig:parallax_vs_motion} we derived the following provisional empirical relation:

\begin{equation}
\begin{split}
    \pi_{\rm{empirical}} = & 0.558657 + 0.0240032\mu_{\rm{total}} \\ + & 0.00157145\mu_{\rm{total}}^2  
    - 3.68780\times10^{-5}\mu_{\rm{total}}^3 \\ + & 3.39325\times10^{-7}\mu_{\rm{total}}^4
\end{split}
\label{eq:prov_pi_vs_mu}
\end{equation}

A drawback of equation \ref{eq:prov_pi_vs_mu} is that it is based on a mixture of luminosity classes and spectral types in the sample of 3,367 stars. In order to improve it, we first obtained the colors of stars in the $r$ and $i^{+}$ Subaru Suprime-Cam bandpasses (centered at wavelengths of 6,288 and 7,683.9 \AA, respectively) from the COSMOS 2015 catalog \citep{COSMOS2015}. These bandpasses were chosen because they were obtained in a homogeneous survey, and their depths are comparable (26.5 and 26.2 mag, respectively). The color excess $E(B - V)$ for all stars that we considered in the COSMOS field was obtained from \citet{COSMOS2015} and was in the range 0.0158--0.0243 mags. The extinction law was taken as 2.660$E(B - V)$ in the $r$ filter and 1.991$E(B - V)$ in the $i^{+}$ filter \citep[][and references therein]{COSMOS2015}, and was applied to de-redden all of the stars in our sample. We generated synthetic $r-i^{+}$ colors of main-sequence, giant, and supergiant stars by convolving either empirical templates of main-sequence and giant stellar spectra \citep{KESSELI2017}, or Kurucz-Lejeune model spectra of supergiants \citep{KURUCZ-LEJEUNE}, with the transmission functions of the $r$ and $i^{+}$ filters from \citet{TANIGUCHI2007}, \citet{COSMOS20}, and \citet{COSMOS2015}. We then compared the observed $r-i^{+}$ colors of the stars with the above synthetic colors, to assign a spectral type to each.

\begin{figure}[htb]
\includegraphics[angle=0,width=3.2in]{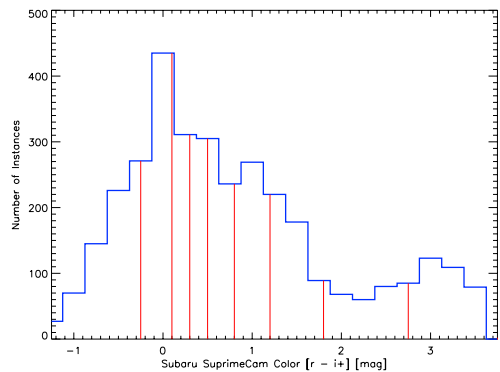}
\caption{The {\it blue line} is the histogram of Subaru Suprime-Cam $r-i^{+}$ colors of 3,386 Gaia DR2 - ACS matches, in the color interval -1.25 $ < r-i^{+} < $ 3.75, which encompasses normal stars. The {\it vertical red lines} denote the boundaries of the $r-i^{+}$ color groups we used in Section \ref{sec:stellar_distances_velocities} as proxy for narrow ranges of spectral types (listed in Table \ref{tab:group_color_fits_Gaia}).}
\label{fig:hist_r_ip_colors_gaia}
\end{figure}

We established relations between Gaia DR2 parallax and total proper motion, analogous to equation \ref{eq:prov_pi_vs_mu}, but restricting stars in the sample within narrow ranges of spectral type and luminosity class. The reason for selecting these narrow ranges was to minimize the effects of varying intrinsic brightness on inferred parallaxes when later applying these relations to non-Gaia stars. Early-type, young stars are expected to be located at smaller Galactic scale heights than late-type, older stars. However, excursions in the Galactic space velocities of stars, and therefore on their proper motions, are expected because of variations in the progenitor initial velocities even among stars formed at common locations, and later on during the life cycles of stars because of stellar encounters. These effects introduce large errors when applying these relations to individual sources. Caution is a must if trying to obtain individual source empirical parallaxes when applying the relations to non-Gaia stars. The relations are instead meant to obtain mean proper motions of ensembles of stars, as described in this section.

\begin{figure*}[htb]
\centering
\includegraphics[angle=0, width=5.7in]{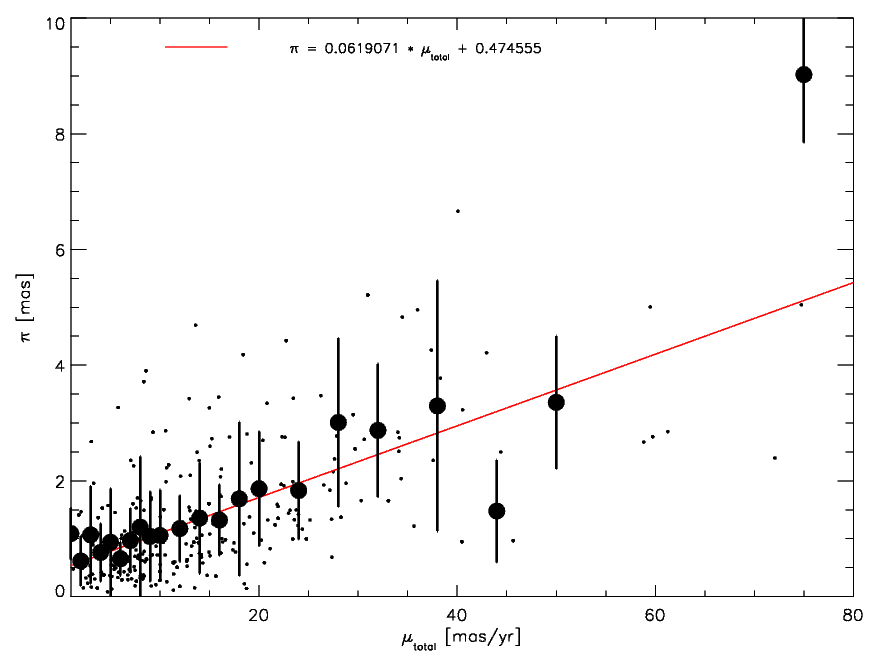}
\caption{Gaia DR2 catalog parallax as a function of total proper motion for early M-type main sequence Gaia DR2-ACS matched stars (whose luminosity class and spectral type were determined as described in the text, with colors in the interval 0.8 $< r-i^{+} <$ 1.2), and with parallaxes $>$ 0. The {\it small black symbols} correspond to individual stars. The {\it large symbols with error bars} are the proper-motion-binned means and standard deviations of parallax. The {\it solid red line} is a fit to the proper-motion-binned means.}
\label{fig:pi_vs_mu_M_dwarfs}
\end{figure*}

We selected Gaia DR2 stars within the color intervals shown in Figure \ref{fig:hist_r_ip_colors_gaia}, which is a histogram of Gaia-ACS matches in these intervals. The photometric uncertainty in each of the $r$ and $i^{+}$ bands was required to be $<$ 20\% in the COSMOS 2015 survey \citep{COSMOS2015}. 

For each Gaia - ACS match with positive Gaia DR2 parallax, we first computed its absolute I-band magnitude assuming each of the luminosity classes (main-sequence, giant, or supergiant), via the distance modulus obtained from its Gaia parallax and SExtractor ACS F814W magnitude. The absolute magnitudes $M_{I}$ as a function of spectral type were obtained from the on-line tabular compilation\footnote{See http://www.pas.rochester.edu/$\sim$emamajek} \citep[referred to by][]{PECAUT2013} for main-sequence stars; from \citet{MIKAMI1982} for giant stars; and from \citet{KEENAN1985} and \citet{MARTINS2005} for supergiant stars. Solar metallicity was assumed in the calculation of $M_{I}$. The luminosity class that yielded the closest match to the apparent magnitude in the F814W ACS bandpass was selected as the luminosity class for the star. Then, stars were grouped by $r-i^{+}$ color and by main-sequence, giant, and supergiant luminosity class, in order to plot and fit a linear relation (or a slow-varying quadratic in a few cases) to the Gaia parallax vs. total proper motion relation for constant luminosity class and within a narrow range of spectral types. Figure \ref{fig:pi_vs_mu_M_dwarfs} shows the Gaia DR2 parallax vs. total proper motion relation for main-sequence stars of color 0.8 $< r-i^{+} <$ 1.2 mag or, per Table \ref{tab:group_color_fits_Gaia}, of spectral type M2--M3. Similar relations of parallax as a function of total proper motion were established for main-sequence and giant stars, in groups of $r-i^{+}$ colors or approximate spectral types listed in Table \ref{tab:group_color_fits_Gaia} (whenever there were sufficient stars). This table also lists the linear or, in a few cases, quadratic fit parameters for the above relations. There were not sufficient supergiant stars to establish a relation. 
For comparison, we also obtained relations between Gaia DR2 parallax and total proper motion, analogous to those in Table \ref{tab:group_color_fits_Gaia}, but for all Gaia DR2 parallaxes, regardless of whether they were positive or negative. We find that the y-intercept $a_{0}$ of the relations is typically $\sim$20--65\% of those in Table \ref{tab:group_color_fits_Gaia}, the slope is $\sim$10--48\% steeper, and the uncertainties in the parameters $a_{0}$, $a_{1}$, and $a_{2}$ are $\sim$10--20\% larger than in Table \ref{tab:group_color_fits_Gaia}. Later on in this section, we quantify the effect of applying these relations to the statistics of proper motions of non-Gaia stars.

\begin{deluxetable*}{ccccc}
\tabletypesize{\footnotesize}
\tablewidth{6.5in}
\tablecolumns{5}
\tablecaption{\sc Polynomial Fits To the Relation of Gaia DR2 Parallax As A Function of Total Proper Motion, Color, and Luminosity Class \tablenotemark{a}}
\tablehead{
\colhead{\sc Color}   & \colhead{\sc Spectral Type} & \colhead{a0} & \colhead{a1} & \colhead{a2}\\
\colhead{$r - i^{+}$} & \colhead{}                  & \colhead{}   & \colhead{}            & \colhead{} }
\startdata 
\multicolumn{5}{l}{\sc Main Sequence Stars} \\
$-$1.25 to $-$0.25 & O4 to B6 &  0.446$\pm$0.032 &  0.0418$\pm$0.0028  & \nodata \\
$-$0.25 to 0.10    & B7 to G0 &  0.708$\pm$0.042 &  0.0325$\pm$0.0034  & \nodata \\
0.10 to  0.30     & G1 to K3 &  0.640$\pm$0.047 &  0.0406$\pm$0.0040  & \nodata \\
0.30 to  0.50     & K4 to K7 &  1.012$\pm$0.139 &  0.0273$\pm$0.0052  & \nodata \\
0.50 to  0.80     & K8 to M1 &  0.481$\pm$0.143 &  0.0585$\pm$0.0054  & \nodata \\
0.80 to  1.20     & M2 to M3 &  0.474$\pm$0.048 &  0.0619$\pm$0.0034  & \nodata \\
1.20 to  1.80     & M4 to M5 &  0.502$\pm$0.151 &  0.0541$\pm$0.0058  & \nodata \\
1.80 to  2.75     & M6 to M8 &  0.891$\pm$0.182 & $-$0.00645$\pm$0.0157 & 0.00184$\pm$0.00023 \\
2.75 to  3
 .75     & M8 to L0 &  1.064$\pm$0.206 & $-$0.0053$\pm$0.0236 & 0.00062$\pm$0.00048 \\
\multicolumn{5}{l}{\sc Giant Stars} \\
0.50 to  0.80     & K8 to M1 & 0.0102$\pm$0.391   & 0.00699$\pm$0.00531  & \nodata \\
0.80 to  1.20     & M2 to M3 & 0.0209$\pm$0.0958  & 0.0051$\pm$0.0102  & \nodata \\
1.20 to  1.80     & M4 to M5 & 0.0559$\pm$0.0154  & 0.0027$\pm$0.0030  & \nodata \\
1.80 to  2.75     & mid to late M      & 0.150$\pm$0.099  & 0.0194$\pm$0.0097 & \nodata \\
2.75 to  3.75     & late M   & 0.180$\pm$0.026    & 0.0040$\pm$0.0039   & \nodata \\
\enddata
\tablenotetext{a}{Polynomial fits to the relation of Gaia DR2 parallax as a function of total proper motion, also as a function of Subaru Suprime-Cam color $r-i^{+}$ and luminosity class (main-sequence or giant stars). These relations were established from 2,703 stars with positive Gaia DR2 parallax, colors in the interval -1.25 $< r-i^{+} <$ 3.75 mag, and photometric uncertainties $<$ 20\% in the $r$ and $i^{+}$ bandpasses. The relations are: $\pi = a_{0} + a_{1}\mu_{total}$, if only $a_{0}$ and $a_{1}$ are listed, and $\pi = a_{0} + a_{1}\mu_{total} + a_{2}\mu_{total}^{2}$ if $a_{0}$, $a_{1}$, and $a_{2}$ are listed. The quoted errors in the linear or quadratic fit coefficients are $\pm$ 1 standard deviation. }
\label{tab:group_color_fits_Gaia}
\end{deluxetable*}

We attempted to derive parallaxes among the 11,519 empirically selected faint stars (matched in ACS and HSC). Of these, 10,818 had COSMOS 2015 Subaru Suprime-Cam colors in the interval -1.25 $< r-i^{+} <$ 3.75 mag and photometric uncertainties $<$ 20\% in the $r$ and $i^{+}$ bandpasses). The relations of Table \ref{tab:group_color_fits_Gaia} were separately applied for main-sequence, giant, and supergiant stars, since we did not know a priori the luminosity class of these sources. The distance modulus as well as absolute magnitudes in the I-band were computed for each of these three luminosity classes, from the corresponding empirical parallax. The luminosity class was identified by finding the closest match of apparent magnitude in the HSC-i band and that from the distance modulus. Among these sources, we determined that 10,816 are main sequence stars, and 2 are giant stars. No supergiants were identified in the sample. With the identified luminosity class and corresponding parallax for each of these sources, we obtained and subtracted the spurious component of proper motion due to the Solar motion relative to the LSR. 

\begin{figure*}[htbp]
\centering
\includegraphics[angle=0, width=7.0in]{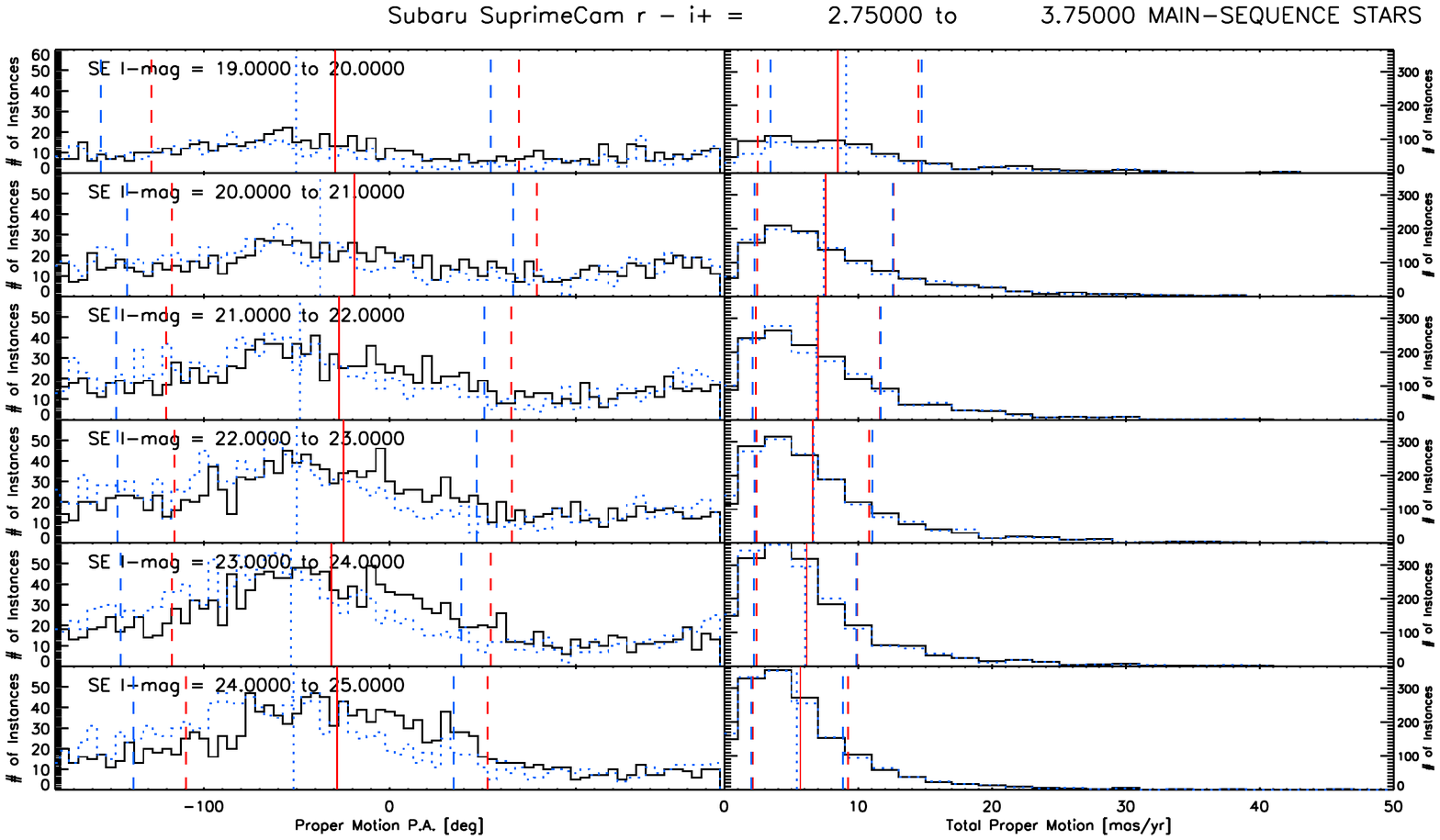}
\caption{The {\it solid black lines} are histograms of proper motion vectors of 8,506 late-type (M8--9) main sequence stars with Subaru SuprimeCam $r-i^{+}$ colors in the range 2.75--3.75 mag and SExtractor AB 19--25 mag. The {\it left panels} are histograms of so-called ``position angle,'' defined here as 0$\arcdeg$ due east, 90$\arcdeg$ due north, and $-90\arcdeg$ due south. The {\it right panels} are histograms of magnitude (quadrature combination of R.A. and Dec. components). Each {\it row} corresponds to a bin of SExtractor AB mag. The {\it red solid vertical line} in each bin is the 3-$\sigma$-clipped mean and the {\it red dashed vertical lines} denote the $\pm$3-$\sigma$-clipped standard deviation. For comparison with the above, using an astrometric grid of all Gaia stars regardless of whether their parallaxes were positive or negative, the {\it blue dotted lines} are analogous histograms, and the {\it blue vertical dotted line} and {\it blue vertical dashed lines} denote analogous 3-$\sigma$-clipped mean and $\pm$3-$\sigma$-clipped standard deviation, respectively.}
\label{fig:hist_pm_vect_mag_pa_MS_lateMtype}
\end{figure*}

After applying the correction of proper motions for the effect of the intrinsic motion of the Solar System, described above, we obtained the proper motions of 10,818 empirically selected faint stars and 1,010 Gaia-ACS stars exclusive of the empirically selected ones. In Figure \ref{fig:hist_pm_vect_mag_pa_MS_lateMtype} we show histograms of the proper motions of 8,506 late-type main-sequence stars, comprised of 8,358 empirically selected faint stars and 148 Gaia-ACS stars, with $r-i^{+}$ colors in the range 2.75--3.75 mag, and with apparent AB mag between 19 and 25. The empirically-derived parallaxes of these sources are in the range 1.05--2.37 mas with a mean of 1.09 mas or a range of distances 421--950 pc with a mean of 915 pc. The median proper motion of these sources is predominantly in the ESE direction, or at a so-called ``position angle'' of $-30\arcdeg$ as shown in Figure \ref{fig:hist_pm_vect_mag_pa_MS_lateMtype}, and defined as 0$\arcdeg$ due east and 90$\arcdeg$ due north. Figure \ref{fig:hist_pm_vect_mag_pa_MS_lateMtype} also shows, as {\it dotted blue lines}, similar histograms of ``position angle'' and vector magnitude of proper motion of the same sources, but after using all Gaia DR2 sources for astrometric correction and empirical parallax estimation, regardless of whether their parallaxes were positive or negative. The histograms are very similar to those using only positive Gaia parallaxes; the mean ``position angle'' is $-50\arcdeg$ and does not alter our conclusions. The mean vector magnitude of proper motion is essentially unchanged. Table \ref{tab:mean_pm_sig_empFS_vs_color_class_mag} lists the mean and standard deviation of proper motions for the 10,818 empirically selected faint main-sequence and giant stars, in bins of $r-i^{+}$ color and SExtractor ACS F814W magnitude. The bins listed in Table \ref{tab:mean_pm_sig_empFS_vs_color_class_mag} were meant to contain stars of not only narrow ranges in spectral type and constant luminosity class, but also narrow ranges in apparent brightness. These bins do not necessarily translate into bins of constant parallax or distance to the Earth, due to dispersion in individual stellar velocities, as explained earlier in this Section. However, the mean distance to Earth was expected to follow an increasing trend for fainter bins within each spectral type and luminosity class bin. Our objective was to probe any pattern in mean proper motions in the COSMOS field, as a function of distance to the Earth. Table \ref{tab:mean_pm_sig_empFS_vs_color_class_mag} lets us identify bins with statistically significant median proper motions, such as those of late-type main-sequence stars plotted in Figure \ref{fig:hist_pm_vect_mag_pa_MS_lateMtype}. Figure \ref{fig:hist_pm_vect_mag_pa_MS_lateMtype} shows that the mean proper motion varies from $\sim$5.8 mas/yr at the farthest distance of $\sim$950 pc to $\sim$9 mas/yr at the closest distance of $\sim$421 pc. Equivalently, Figure \ref{fig:hist_pm_vect_mag_pa_MS_lateMtype} also shows that the fraction of stars with proper motions $>10$ mas/yr out of the total number of stars in each bin varies from 16\% at the farthest distance to 37\% at the closest distance. 

\begin{deluxetable*}{ccccccc}
\tabletypesize{\footnotesize}
\tablewidth{6.5in}
\tablecolumns{7}
\tablecaption{\sc Proper Motions in R.A. and Dec. As A Function of Luminosity Class, Color, and SExtractor AB Magnitude\tablenotemark{a}}
\tablehead{
\colhead{\sc Color} & \multicolumn{6}{c}{\sc SExtractor AB Magnitude} \\
\colhead{$r - i^{+}$} & \colhead{19--20} & \colhead{20--21} & \colhead{21--22} & \colhead{22--23} & \colhead{23-24} & \colhead{24-25} }
\startdata 
\multicolumn{7}{c}{\sc Main Sequence Stars} \\
$-$1.25 to $-$0.25 &  0.77$\pm$1.41 & -0.96$\pm$1.07 & -1.57$\pm$0.71 &  0.95$\pm$0.63 &  0.74$\pm$0.67 &  1.68$\pm$0.70 \\
                   & -3.33$\pm$1.57 & -1.02$\pm$0.56 & -1.02$\pm$0.55 & -1.98$\pm$0.56 & -1.82$\pm$0.76 & -1.22$\pm$0.81 \\
$-$0.25 to $+$0.10 &  2.00$\pm$2.60 & -1.96$\pm$1.91 &  0.41$\pm$0.86 &  0.99$\pm$0.66 &  1.06$\pm$0.67 &  2.11$\pm$0.66 \\
                   & -0.64$\pm$2.92 & -4.30$\pm$1.35 & -3.05$\pm$0.86 & -1.74$\pm$0.78 & -0.65$\pm$0.70 & -1.34$\pm$0.50 \\
$+$0.10 to $+$0.30 &  2.11$\pm$2.57 & -1.56$\pm$1.19 & -0.65$\pm$0.97 &  1.25$\pm$1.01 &  0.97$\pm$0.71 &  1.33$\pm$0.79 \\
                   & -5.59$\pm$1.58 & -2.52$\pm$1.11 & -2.45$\pm$0.76 & -1.52$\pm$0.78 & -1.52$\pm$0.77 & -1.64$\pm$0.82 \\
$+$0.30 to $+$0.50 & -0.15$\pm$1.32 &  1.42$\pm$1.25 &  2.20$\pm$1.22 &  0.93$\pm$1.06 &  1.47$\pm$0.74 &  3.89$\pm$0.63 \\
                   & -3.67$\pm$1.47 & -1.06$\pm$0.94 & -0.46$\pm$1.09 & -1.21$\pm$0.81 & -1.50$\pm$0.64 & -1.64$\pm$0.88 \\
$+$0.50 to $+$0.80 & -0.12$\pm$1.78 & -1.12$\pm$1.30 &  1.09$\pm$0.82 &  0.87$\pm$0.94 &  0.77$\pm$0.63 &  2.34$\pm$0.69 \\
                   & -2.42$\pm$1.86 & -3.52$\pm$0.98 & -2.17$\pm$0.77 & -1.16$\pm$0.83 & -2.11$\pm$0.46 & -1.26$\pm$0.61 \\
$+$0.80 to $+$1.20 &  1.30$\pm$1.94 &  1.29$\pm$1.61 &  1.79$\pm$0.69 & -0.05$\pm$0.98 &  0.41$\pm$0.57 &  2.24$\pm$0.59 \\
                   & -0.17$\pm$1.64 & -1.05$\pm$1.25 & -0.55$\pm$0.87 & -2.02$\pm$0.97 & -2.62$\pm$0.63 & -0.63$\pm$0.74 \\
$+$1.20 to $+$1.80 & -4.01$\pm$0.88 & -0.11$\pm$1.01 & -0.40$\pm$0.61 &  1.97$\pm$1.07 &  1.53$\pm$0.60 &  2.02$\pm$0.54 \\
                   & -2.59$\pm$1.29 & -0.17$\pm$0.99 & -1.69$\pm$0.67 & -3.48$\pm$0.94 & -1.93$\pm$0.40 & -2.06$\pm$0.54 \\
$+$1.80 to $+$2.75 & -3.27$\pm$2.96 &  2.81$\pm$1.43 &  3.39$\pm$1.13 &  2.22$\pm$0.74 &  2.37$\pm$0.71 &  1.82$\pm$0.61 \\
                   & -0.04$\pm$2.09 & -3.19$\pm$1.39 & -0.44$\pm$0.69 & -1.46$\pm$1.06 & -0.82$\pm$0.52 & -0.66$\pm$0.66 \\
$+$2.75 to $+$3.75 &  0.87$\pm$0.33 &  1.32$\pm$0.22 &  1.64$\pm$0.17 &  1.55$\pm$0.16 &  2.13$\pm$0.14 &  2.25$\pm$0.13 \\
                   & -1.93$\pm$0.30 & -0.92$\pm$0.21 & -1.57$\pm$0.17 & -1.36$\pm$0.16 & -1.50$\pm$0.13 & -1.34$\pm$0.12 \\
\multicolumn{7}{c}{\sc Giant Stars} \\
$+$0.50 to $+$0.80 & \nodata        & \nodata        &  0.39$\pm$2.55 & \nodata        & \nodata        & -0.28$\pm$1.67 \\
                   & \nodata        & \nodata        &  0.07$\pm$1.60 & \nodata        & \nodata        & -0.48$\pm$1.97 \\
\enddata
\tablenotetext{a}{Each table entry corresponds to a given $r-i^{+}$ color range (``row'') and SExtractor AB mag range (``column''), separately for main-sequence and giant stars. In the case of main-sequence stars, each table entry lists median $\pm$ 3$\sigma$-clipped standard deviation of the mean of proper motion in R.A. (upper row within table entry) and Dec. (lower row within table entry). In the case of giant stars, only two objects were identified in our sample of faint stars and each belongs in a separate SExtractor AB magnitude range so their proper motions are for individual objects and their uncertainties are as calculated in Section \ref{sec:uncertainties}.}
\label{tab:mean_pm_sig_empFS_vs_color_class_mag}
\end{deluxetable*}

At the mean Galactic coordinates of the COSMOS field (236.6$\arcdeg$,$+$42.1$\arcdeg$), the galactocentric Cartesian coordinates $(x,y,z)$ (each in kpc) are, as a function of distance $D$ (kpc) to the Sun:

\begin{equation}
x = -0.406 D -8.2
\end{equation}
\begin{equation}
y = -0.621 D
\end{equation}
\begin{equation}
z = +0.670 D
\end{equation}
where $x$ points from the Sun and is positive towards the Galactic center, $y$ is parallel to the Galactic plane and is positive in the Galactic rotation direction, and $z$ is the vertical distance to the Galactic plane and is positive towards the North Galactic pole, as described by, e.g., \citet{YAN20}. In this coordinate system, the Sun is located at (-8.2, 0, 0.015) (kpc).

Given proper motions $\mu_{\alpha}$ (R.A.) and $\mu_{\delta}$ (Dec.) in mas/yr, radial velocity $v_{rad}$ in km/s (along the line-of-sight by definition), and distance $D$ in kpc, Galactic space velocities $(U,V,W)$ relative to the LSR were calculated using the procedure by \citet{JOHNSON_SODERBLOM}, except that the velocity component $U$, parallel to the Galactic plane and pointing away from the Galactic center, is positive in this last sense. The velocity component $V$ is also parallel to the Galactic plane and points in the direction of Galactic rotation, while $W$ is perpendicular to the Galactic plane and is positive towards the north Galactic pole.

\begin{figure}[htbp]
\centering
\includegraphics[angle=0, width=3.2in]{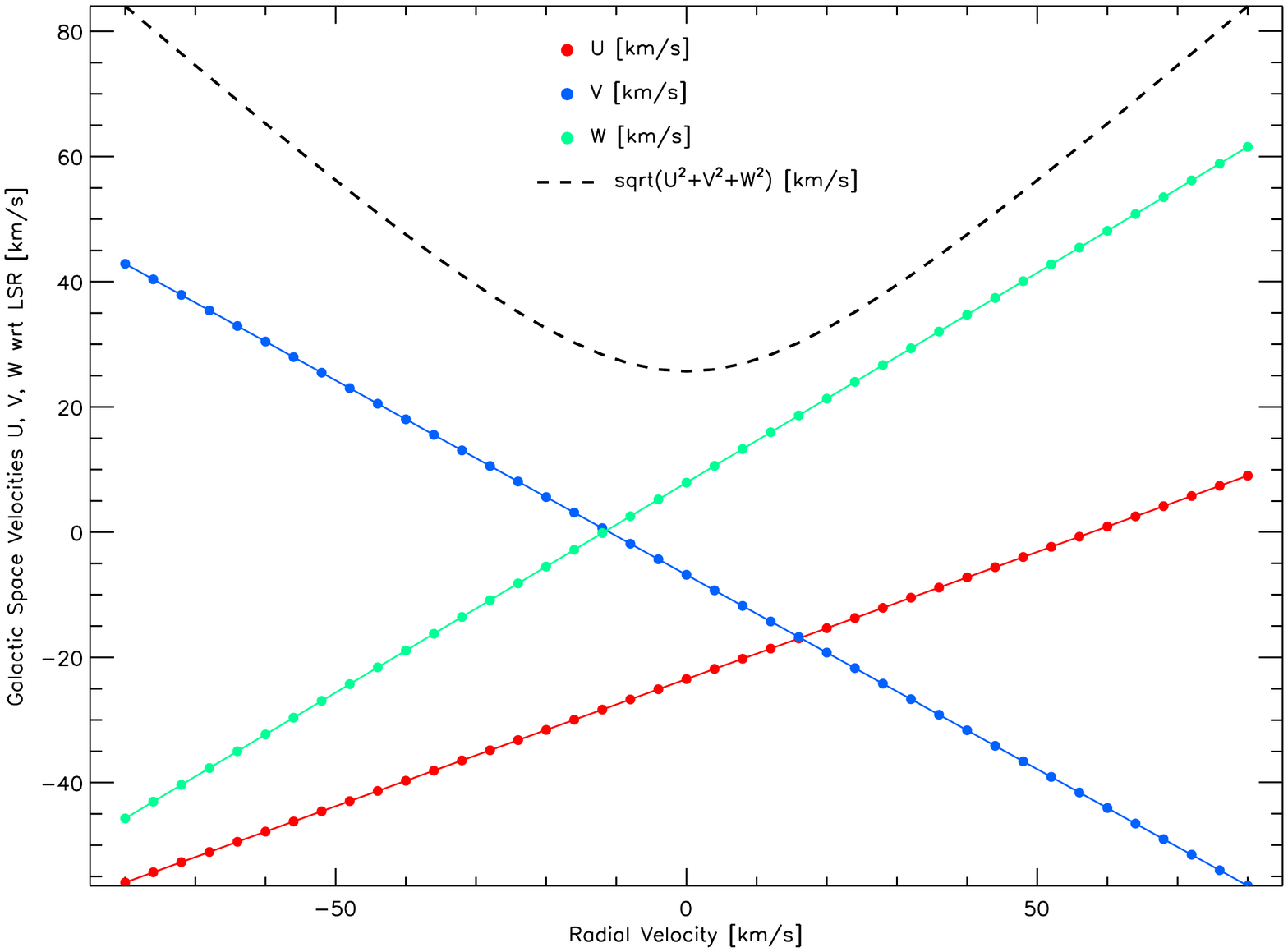}
\caption{Galactic space velocity components $U$, $V$, $W$ for late-type main-sequence stars with Subaru SuprimeCam colors $2.75 < r - i^{+} < 3.75$ and SExtractor F814W AB magnitude in the range 24 to 25. The $(U,V,W)$ velocities were computed from a mean proper motion of 5.8 mas/yr directed $-$30$\arcdeg$ (clockwise) from East, as obtained from 1,540 stars whose histograms of proper motions are shown in the bottom panel of Figure \ref{fig:hist_pm_vect_mag_pa_MS_lateMtype}. The mean parallax for these stars of 1.07 mas was also assumed, and radial velocities were varied from $-$80 km/s to $+$80 km/s. The {\it red symbols} are the $U$ velocities; the {\it blue symbols} are the $V$ velocities; the {\it green symbols} are the $W$ velocities and the {\it black dashed line} denotes the total velocities.}
\label{fig:uvw_velocities_mean_mu_and_pi_faint_MS_lateMtype}
\end{figure}

The faintest late-type main-sequence stars in Figure \ref{fig:hist_pm_vect_mag_pa_MS_lateMtype} (1,540 stars with AB mag in the range 24--25) have mean proper motions ${<}\mu_{\alpha}{>} \sim 5.02$ mas/yr and ${<}\mu_{\delta}{>} \sim -2.9$ mas/yr, as derived from the bottom panels in Figure \ref{fig:hist_pm_vect_mag_pa_MS_lateMtype}, and mean parallax $\sim$1.07 mas. For this group of stars we calculated the $U$, $V$, and $W$ velocity components from their individual proper motions and parallaxes, assuming at first zero radial velocities, to get a range of components from $-$99 km/s to $+$81 km/s. We then used this range to estimate that $v_{rad}$ can be, at most, in the range from $\sim -$80 km/s to $+$80 km/s. We then computed the $U$, $V$, and $W$ velocity components corresponding to the mean proper motion and mean parallax of these stars, and assuming radial velocities in the above range. Figure \ref{fig:uvw_velocities_mean_mu_and_pi_faint_MS_lateMtype} shows the resulting possible velocities corresponding to the mean proper motions and parallax, and radial velocities from $-$80 km/s to $+$80 km/s.

From Figure \ref{fig:uvw_velocities_mean_mu_and_pi_faint_MS_lateMtype} it can be seen that the rotational ($V$) and vertical ($W$) velocities are very nearly equal in magnitude and of opposite sign for any given radial velocity. They are $\sim$0 km/s for a radial velocity of $\sim -$12 km/s. The velocity component $V$ does not dominate over the other components $U$ and $W$ for any radial velocity. Thus the mean motion of these faint late-type main sequence stars is not predominantly along the Galactic rotation. It is unlikely that vertical velocity $W$ is large, which would place limits on extreme radial velocity. If $-10$ km/s $< W < +$10 km/s then $-30$ km/s $< v_{rad} < 0$ km/s. The $U$ velocity  is $< 0$ for most radial velocities. That is, the mean or predominant motion of late-type main-sequence stars at $\sim$934 pc is towards the Galactic center. In the range of radial velocities $-30$ km/s $< v_{rad} < 0$ km/s, $U$ dominates over both $V$ and $W$ by a factor of at least $\sim$2. 

\begin{figure*}
\centering
\includegraphics[angle=0, height=6in]{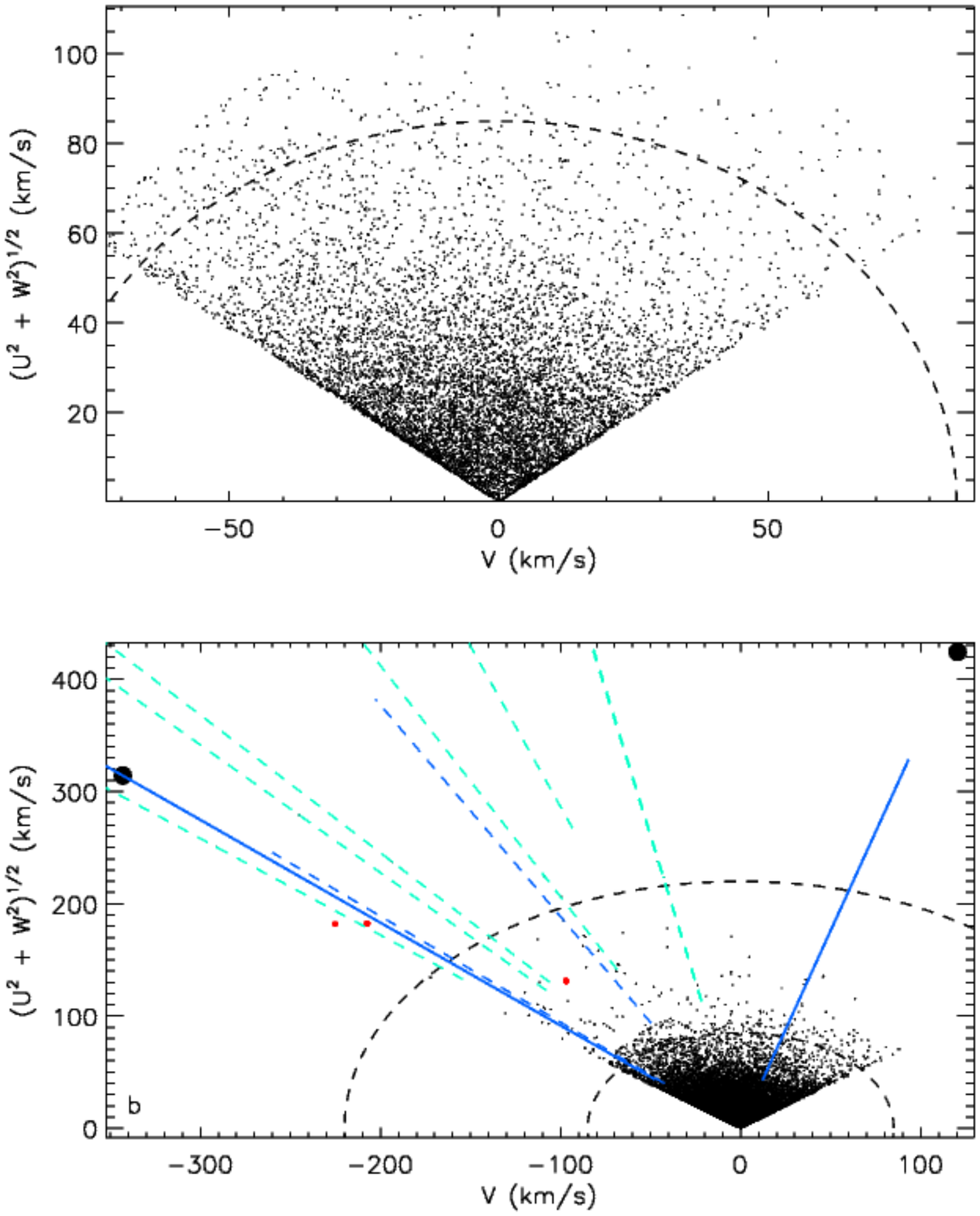}
\caption{{\it (a) Upper panel:} Toomre diagram of 8,506 late-type main-sequence stars, comprised of 8,358 empirically selected stars and 148 Gaia-ACS matches, with Subaru SuprimeCam colors $2.75 < r - i^{+} < 3.75$ and SExtractor F814W AB magnitude in the range 19 to 25. The {\it dashed line} denotes total velocity $v_{tot} = 85$ km/s separating the thin and thick disk components of the galaxy. {\it Dots} denote the velocity components of each source. {\it (b) Lower panel:} Toomre diagram of 11,830 stars, comprised of 10,818 from our sample of empirically selected stars, 1,010 from Gaia DR2-ACS matches with Subaru SuprimeCam colors $-1.25 < r - i^{+} < 3.75$ and SExtractor F814W AB magnitude in the range 19 to 25, of which the sources in (a) are a subset, and 2 Gaia DR2-ACS matches as above, but with SExtractor F814W AB mag $<$ 18.5. The {\it black dashed} lines represent $v_{tot}$ of 220 km/s and 85 km/s, to delimit the loci of the halo, the thick and thin disks respectively. The {\it large filled circles} denote initial estimates of the space velocities of two stars, listed in the first 2 rows of Table \ref{tab:halo_stars}, assuming solar metallicity giant star absolute magnitudes, that we estimate to be candidate members of the halo. The {\it blue solid lines} are re-derived space velocities of these two stars assuming metal-poor giant star absolute magnitudes, for consistency with the halo, as explained in the text, and also using Monte Carlo simulations in view of large parallax uncertainties, described in Appendix \ref{sec:appendix_B}, yielding distance and corresponding velocity ranges for each star. The {\it small red filled circles} are re-derived space velocities of three metal-poor giant stars, assumed to be at a distance of 20 kpc, and listed in rows 3--5 of Table \ref{tab:halo_stars}. The {\it blue and green dashed lines} are ranges of space velocities for Gaia-ACS matched sources, consisting respectively of two bright (SExtractor F814W AB mag $<$ 18.5) giant stars from the sample in the {\it upper-left panel} of Figure \ref{fig:pi_vs_mu_KM_giants_cosmos_and_highlat}, and six Gaia stars not in the sample of empirically selected faint stars. The two bright giant stars are at distances possibly consistent with 20 kpc at the Sangarius stream, but they are inconsistent with the metal-poor environment in this stream. They are listed in rows 6-7 of Table \ref{tab:halo_stars}. The six Gaia stars not in the empirical faint star sample are candidate high velocity stars but at relatively near distances, as explained in the text. The parallaxes of the above eight sources from Gaia DR2 had large relative uncertainties and therefore we derived ranges of distance and corresponding velocity using Monte Carlo simulations analogous to those in Appendix \ref{sec:appendix_B}; see also  \citet{GAIADR2PARALLAX}.}
\label{fig:toomre_diags}
\end{figure*}

In order to estimate membership of the 8,358 empirically selected faint stars and the 148 Gaia-ACS matched stars, of late-type main-sequence classification, in the thin and thick disks of the Galaxy, we calculated lower limits (because of the absence of radial velocity information) to $(U,V,W)$ space velocities relative to the LSR for each of them. We then plotted a Toomre diagram, consisting of $\sqrt{U^{2} + W^{2}}$ as a function of $V$ \citep{TOOMRE,YAN20}. We distinguished stars in the thin disk, thick disk, and halo based on whether the total velocity $v_{tot} = \sqrt{U^{2} + V^{2} + W^{2}}$ is, respectively, $v_{tot} < 85$ km/s, 85 km/s $< v_{tot} <220$ km/s, and $v_{tot} >220$ km/s \citep{VENN04,NISSEN04,YAN20}. The Toomre diagram of these stars is shown in Figure \ref{fig:toomre_diags}a, where it can be seen that the majority of these stars are in the thin disk. The delimited ``fan''-like appearance of the locus of these stars is due to the fact that radial velocity $v_{rad}$ has been assumed to be zero. The inclusion of non-zero $v_{rad}$ would not alter our conclusion that most stars have $v_{tot} < 85$ km/s for most values of $v_{rad}$, as suggested by the {\it dashed line} in Figure \ref{fig:uvw_velocities_mean_mu_and_pi_faint_MS_lateMtype}, and the fact that there are much fewer stars with $v_{tot}$ at or near the thin disk limit of 85 km/s in Figure \ref{fig:toomre_diags}a.

We next searched for any candidate high-velocity stars, belonging to the halo, in the whole sample of stars with JSP proper motions in the COSMOS field. The sample of late-type main-sequence stars plotted in Figure \ref{fig:toomre_diags}a does not include any such halo members. Figure \ref{fig:toomre_diags}b is the Toomre diagram of all 11,828 stars of which 10,818 are from our empirically selected faint stars sample and 1,010 are from Gaia-ACS matches with Subaru SuprimeCam colors $-1.25 < r - i^{+} < 3.75$ and with SExtractor AB mag in the F814W ACS filter between 19 and 25, for which empirical parallaxes, spectral types and luminosity classes could be derived. Figure \ref{fig:toomre_diags}b shows that two sources, shown as {\it large black filled circles} are candidate high-velocity stars in the Galactic halo. The first two rows of Table \ref{tab:halo_stars} list these two sources. It should be recalled that the derivation of the empirical parallaxes needed to derive the space velocities in Figures \ref{fig:toomre_diags}a and \ref{fig:toomre_diags}b assumed solar metallicity, as described earlier in this Section. We re-derived the empirical parallaxes and distances to these two sources by assuming metal-poor giant star absolute magnitudes at the Galactic halo. For this purpose we used the Dartmouth Stellar Evolution Program database\footnote{http://stellar.dartmouth.edu/models/grid.html} \citep[DSEP]{DARTMOUTH} and selected an isochrone for an age of 13 Gyr, metallicity of [Fe/H] = -2.0 dex and [$\alpha$/Fe] = -0.2 dex, and bandpasses in the SDSS {\it ugirz} photometric system. The Subaru SuprimeCam $r - i^{+}$ and SDSS $r - i$ colors are very similar, as we confirmed by comparing our synthetic colors of main-sequence and giant stellar spectra by \citet{KESSELI2017} and corresponding SDSS colors for the same spectra (op cit.). By comparing the above metal-poor isochrone with the corresponding solar metallicity one it is seen that metal-poor giants are more luminous than solar metallicity ones. In particular, K8- and M0-type metal-poor giants would be factors of $\sim$6.6 and 6.1, respectively, farther away than solar metallicity ones for constant apparent {\it i}-magnitude. We also re-derived the empirical parallaxes or distances to these two stars by performing Monte Carlo simulations of distance, as explained in Appendix \ref{sec:appendix_B}. The errors in parallax of these distant stars and the Gaia DR2 stars used to derive the relations in Table \ref{tab:group_color_fits_Gaia} for giants are very large, as described in Appendix \ref{sec:appendix_B}. Therefore a straightforward application of the relations for giants in Table \ref{tab:group_color_fits_Gaia}, as carried out to generate the space velocities shown as {\it large black filled circles} in Figure \ref{fig:toomre_diags}b, was deemed inexact. Instead, the application of the Monte Carlo simulations of distance resulted in the median distance and asymmetric upper and lower uncertainties listed in the first two rows of Table \ref{tab:halo_stars}. Corresponding re-derived Galactic space velocities for these two sources are shown in Figure \ref{fig:toomre_diags} as {\it blue solid lines} representing the ranges of distances per the upper and lower uncertainties in Table \ref{tab:halo_stars}. The uncertainties of the space velocities of these two giant stars are large but candidate halo membership is possible.

\begin{deluxetable*}{ccccccccc}
\tabletypesize{\footnotesize}
\rotate
\tablewidth{6.5in}
\tablecolumns{9}
\tablecaption{\sc Candidate high-velocity Stars In the JSP Proper Motion Sample}
\tablehead{
\colhead{\sc R.A. (J2000.0)} & \colhead{\sc Dec. (J2000)} & \multicolumn{2}{c}{\sc JSP Proper Motions\tablenotemark{a}} & \colhead{\sc ACS F814W mag} & \colhead{\sc Color $r - i^{+}$\tablenotemark{b}} & \colhead{\sc Spectral Type\tablenotemark{c}} & \colhead{\sc Luminosity Class\tablenotemark{c}} & \colhead{\sc Distance\tablenotemark{d}}\\
\colhead{} & \colhead{} & \colhead{$\mu_{\alpha}$} & \colhead{$\mu_{\delta}$} & \colhead{} & \colhead{} & \colhead{} & \colhead{} & \colhead{} \\
\colhead{h, m, s} & \colhead{d, m, s} & \colhead{mas/yr} & \colhead{mas/yr} & \colhead{mag [AB]} & \colhead{mag [AB]} & \colhead{} & \colhead{} & \colhead{kpc} }
\startdata 
09 58 09.73415 & $+$02 17 57.0692 &  $+$0.39$\pm$2.55 & $+$0.07$\pm$1.60 & 21.738 & 0.6110$\pm$0.0051 & M0 & III & 54$^{+129}_{-30}$ \\
10 02 30.70664 & $+$01 54 41.5045 &  $-$0.28$\pm$1.67 & $-$0.48$\pm$1.97 & 24.735 & 0.529$\pm$0.018   & K8 & III & 59$^{+156}_{-35}$ \\
\multicolumn{9}{c}{\sc Candidates From Color-Magnitude Diagram\tablenotemark{e}} \\
09 57 58.90735 & $+$02 03 48.5148 &  $+$0.88$\pm$1.93 & $-$1.48$\pm$1.72 & 19.418 & 0.108$\pm$0.016   & F9 & III & \nodata \\
09 58 10.82833 & $+$02 02 21.8468 &  $-$0.79$\pm$0.87 & $-$2.95$\pm$0.92 & 19.056 & 0.1162$\pm$0.0036 & F9 & III & \nodata \\
09 58 18.41377 & $+$02 31 42.3111 &  $+$0.38$\pm$0.73 & $-$2.89$\pm$2.31 & 20.025 & 0.1004$\pm$0.0036 & F9 & III & \nodata \\
\multicolumn{9}{c}{\sc Questionable Sources From Bright Gaia DR2 Solar Metallicity Stars\tablenotemark{f}} \\
09 58 11.57578 & $+$01 58 31.6744 &  $-$1.857$\pm$0.067\tablenotemark{g} & $-$0.950$\pm$0.068\tablenotemark{g} & 15.780 & 0.7224$\pm$0.0078 & M1 & III & 21.2$^{+22.5}_{-10.4}$\tablenotemark{f} \\
10 01 20.02097 & $+$02 01 05.3836 &  $-$1.68$\pm$0.54\tablenotemark{g}   & $-$2.68$\pm$0.99\tablenotemark{g}   & 17.087 & 0.6945$\pm$0.0061 & M1 & III & 8.6$^{+15.3}_{-4.7}$\tablenotemark{f} \\
\enddata
\tablenotetext{a}{Corrected for the effect of Solar motion, as explained in Section \ref{sec:stellar_distances_velocities}.}
\tablenotetext{b}{Colors are from the Subaru Suprime-Cam COSMOS2015 survey \citep{COSMOS2015}.}
\tablenotetext{c}{Spectral types were derived from synthetic colors of main sequence and giant stars using models from \citet{KESSELI2017}, as explained in Section \ref{sec:stellar_distances_velocities}. Luminosity class was inferred from the observed SExtract F814W AB magnitude and absolute magnitudes of main sequence and giant stars (see references in Section \ref{sec:stellar_distances_velocities}.}
\tablenotetext{d}{Starting from relations between Gaia proper motions and parallaxes, as explained in Section \ref{sec:stellar_distances_velocities}, Monte Carlo simulations of variations in these proper motions and parallaxes, and of variations in the JSP proper motion of the source itself, were carried out to derive asymmetric distributions of simulated distance as explained in Appendix \ref{sec:appendix_B}. In addition, the distance moduli used to compute empirical distances to these sources used metal-poor giant star absolute magnitudes, as explained in the text. For candidate sources from a color-magnitude diagram, where no data are shown, distances were assumed to be 20 kpc.}
\tablenotetext{e}{Derived from a color-magnitude diagram for metal poor giant stars at the distance to the Sangarius stream, as explained in Section \ref{sec:discussion}.}
\tablenotetext{f}{Sources are from the sample of K8--M1 giant stars from which a Gaia parallax vs. total proper motion relation was established and applied to other sources (see the {\it upper-left panel} of Figure \ref{fig:pi_vs_mu_KM_giants_cosmos_and_highlat} in Appendix \ref{sec:appendix_B}). In view of large relative uncertainties of Gaia parallaxes, Monte Carlo simulations of parallax were carried out to derive asymmetric distributions of simulated distance, analogous to the simulations in Appendix \ref{sec:appendix_B}; see also \citet{GAIADR2PARALLAX}. The classification of these sources as giants is based on solar metallicity models, explained in Section \ref{sec:stellar_distances_velocities}. }
\tablenotetext{g}{Proper motions from the Gaia DR2 catalog, corrected for the effect of Solar motion.}
\label{tab:halo_stars}
\end{deluxetable*}

We computed distances from Gaia DR2 parallaxes for the five K8--M1 giant stars that were used to generate the relation between Gaia parallax and total proper motion listed in Table \ref{tab:group_color_fits_Gaia}, and whose data are plotted in the {\it upper-left} panel of Figure \ref{fig:pi_vs_mu_KM_giants_cosmos_and_highlat}. That is, we attempted to find out if these sources were themselves candidate high-velocity and/or candidate members of the Sangarius stream. These stars were drawn from among the 3,937 Gaia-ACS matched sources, which are unconstrained in apparent magnitude and therefore were not previously analyzed above because they were brighter than SExtractor F814W 18.5 mag. The relative uncertainties of Gaia parallax for these sources are large and therefore we performed Monte Carlo simulations of parallax, using 500 Gaussian deviates, to estimate median distances and upper and lower asymmetric uncertainties, in analogy to Appendix \ref{sec:appendix_B}; see \citet{GAIADR2PARALLAX}. Of these five K8--M1 giant stars, three had distances smaller than 4 kpc and were discarded from our sample. The remaining two sources are listed in the last two rows of Table \ref{tab:halo_stars} and their space velocities are shown as statistical ranges in Figure \ref{fig:toomre_diags}b as {\it blue dashed lines}. 

Figure \ref{fig:toomre_diags}b shows that six stars from the sample of 1,010 Gaia-ACS matches exclusive of empirically selected faint stars, are candidate high-velocity objects. The relative uncertainties of Gaia DR2 parallax were $>0.2$ for these objects. Therefore, to estimate their distances we performed Monte Carlo simulations analogous to that shown in Appendix \ref{sec:appendix_B}, for consistency with \citet{GAIADR2PARALLAX}. Table \ref{tab:gaia_high_vel} in Appendix \ref{sec:appendix_C} lists the coordinates, median distances, and upper and lower asymmetric uncertainties of distances of these six stars. It can be seen that their distances range from $\sim$480 pc to $\sim$6,170 pc and therefore that they are relatively nearby. These candidate high velocity stars are not further analyzed here but the reader is referred to Table \ref{tab:gaia_high_vel} for follow-up.

\section{Discussion} \label{sec:discussion}

\begin{figure}
\hspace{-0.5in}
\includegraphics[angle=0, width=4.2in]{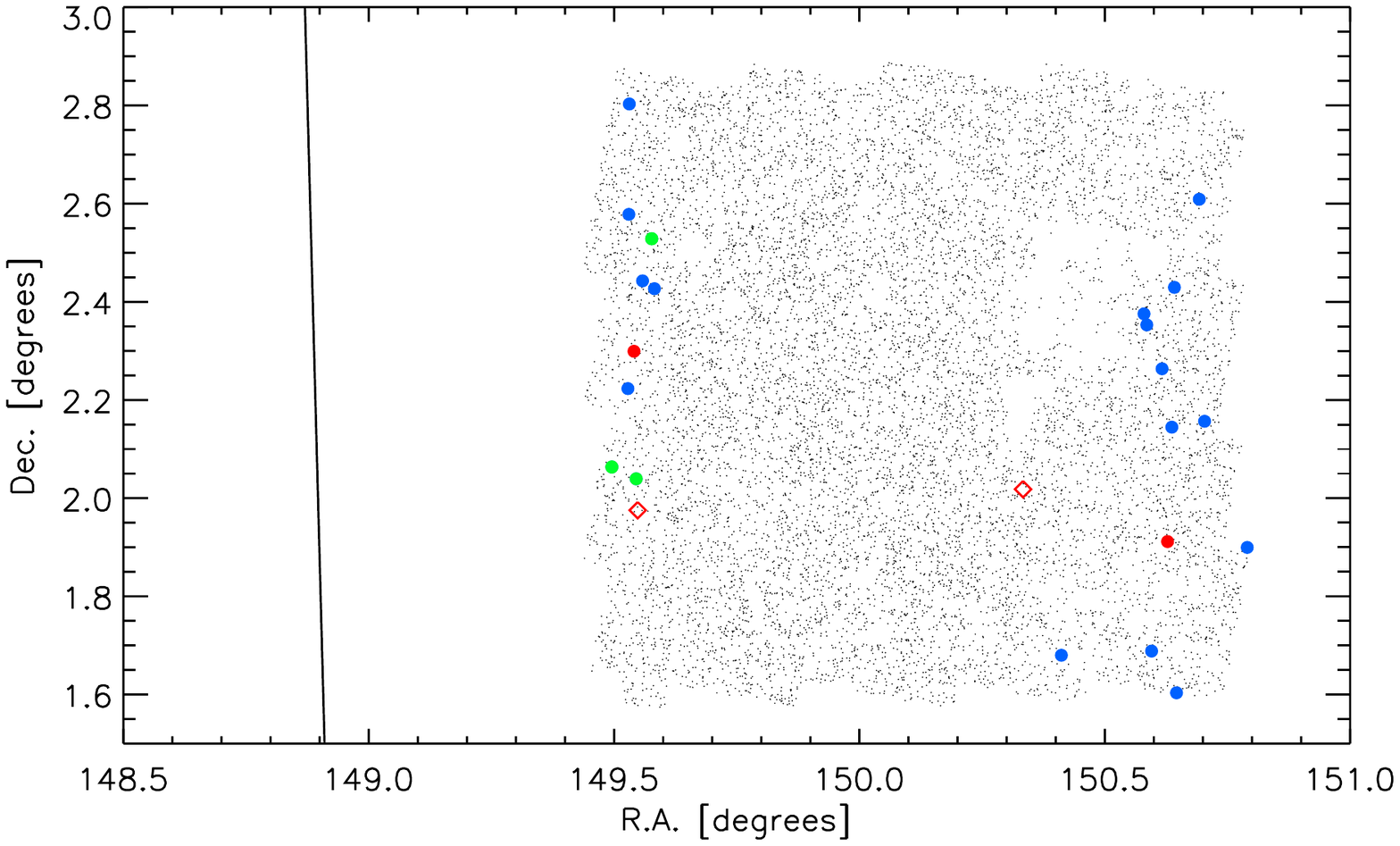}
\caption{The {\it dots} are the coordinates of the sample of 10,818 stars with JSP proper motions and with Subaru SuprimeCam colors $-1.25 < r - i^{+} < 3.75$ and SExtractor F814W AB magnitude in the range 19 to 25. The {\it solid line} delineates the center of the Sangarius stream \citep{GRILLMAIR17}. The {\it colored filled circles} are the coordinates of 21 candidate Sangarius stream stars in our sample, of which {\it red circles} are the two candidate high-velocity stars in the Toomre diagram in Figure \ref{fig:toomre_diags}b listed in the first two rows of Table \ref{tab:halo_stars}, and {\it blue and green circles} are the coordinates of 19 sources obtained from a color-magnitude diagram of metal poor giant stars at the distance to the Sangarius stream. Of these 19 stars, the 3 stars whose coordinates are shown as {\it green filled circles} have proper motions consistent with the expected motions in the Sangarius stream, as further discussed in this Section and in Figure \ref{fig:expected_sangarius_proper_motions}. The {\it red open diamonds} are the positions of two bright giant stars at distances similar to the Sangarius stream, but of solar metallicity which is inconsistent with the stellar population of this stream. These two stars are listed in the last two rows of Table \ref{tab:halo_stars} as questionable sources as regards candidate membership in Sangarius.}
\label{fig:coordinates_all_empFS_and_sangarius}
\end{figure}

To investigate whether the two giant stars in the first two rows of Table \ref{tab:halo_stars} are candidate members of the Sangarius stream, which is a dynamically ``cold'' structure located at a distance of $\sim$20 kpc and which extends for tens of degrees with a width of $\sim 1\arcdeg$ \citep{GRILLMAIR17}, we note that the stream passes $\sim 0.5\arcdeg$ west of the COSMOS field as covered by ACS observations (Figure \ref{fig:coordinates_all_empFS_and_sangarius}). It is possible that the eastern edge of the stream passes through the western edge of the COSMOS field. The positions of these two giant stars identified from proper motions and colors $r - i^{+}$ are shown as {\it red filled circles} in Figure \ref{fig:coordinates_all_empFS_and_sangarius}. Of these two stars, the M0 giant is located near the eastern edge of the Sangarius stream and the K8 giant is located near the eastern edge of the COSMOS field. The lower limits of distances to these two sources are $\sim$24 kpc (Table \ref{tab:halo_stars}) and therefore suggest that membership in the Sangarius stream is possible.

\begin{figure}[htbp]
\centering
\includegraphics[width=3.2in]{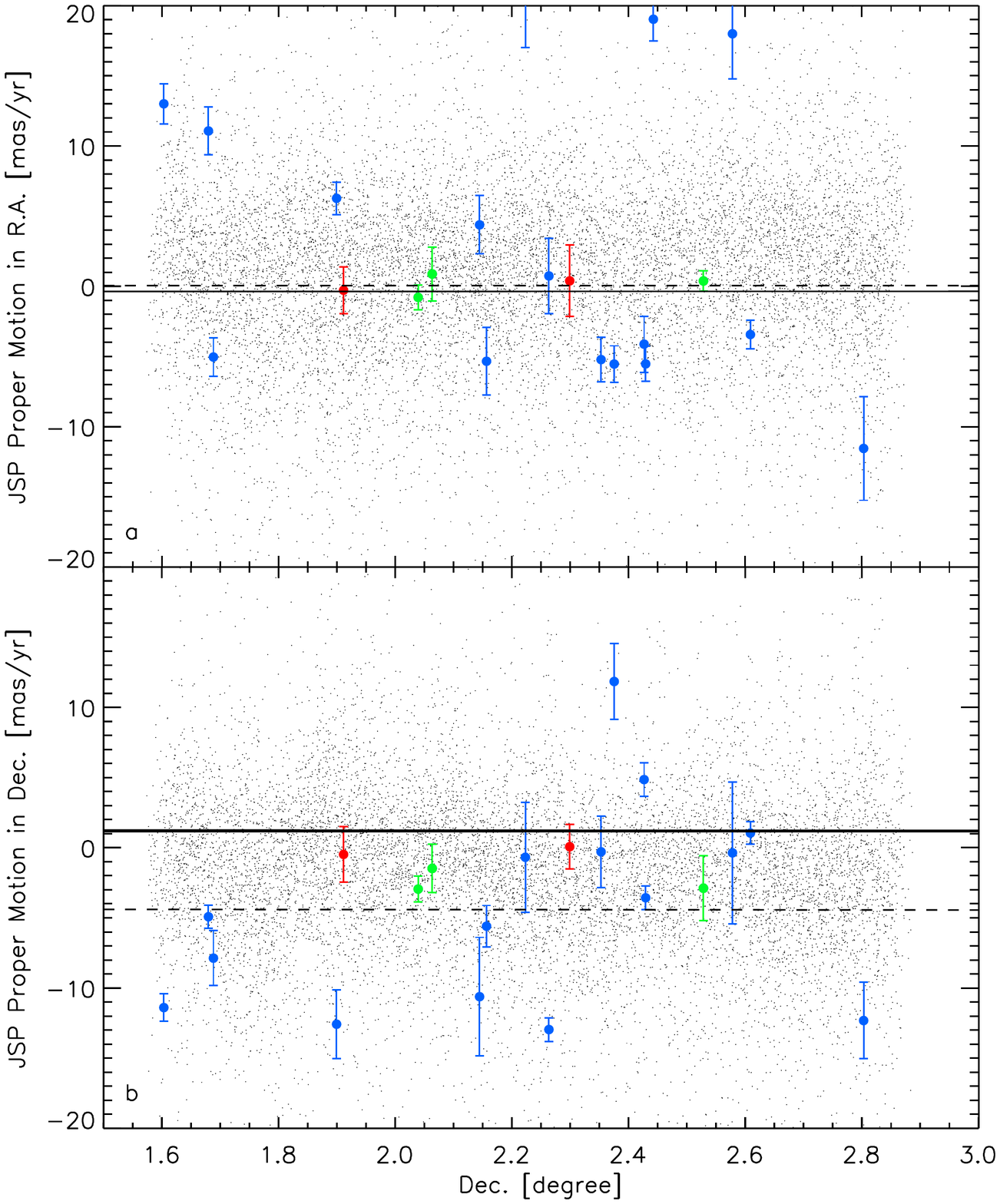}
\caption{{\it (a) Upper panel:} The {\it dots} are the JSP proper motions in R.A., as a function of Dec., for the 10,818 stars with Subaru SuprimeCam colors $-1.25 < r - i^{+} < 3.75$ and SExtractor F814W AB magnitude in the range 19 to 25. The {\it nearly horizontal lines} are the R.A. proper motions as a function of Dec. expected in the Sangarius stream for either prograde ({\it solid line}) or retrograde ({\it dashed line}) orbits (\citep{GRILLMAIR17}. The {\it red filled circles with error bars} are the proper motions of two candidate Sangarius stream sources selected from their high space velocities in the Toomre diagram in Figure \ref{fig:toomre_diags}b, listed as the first two rows in Table \ref{tab:halo_stars}. The {\it blue and green filled circles} are the proper motions of 19 sources obtained from a color-magnitude diagram of metal poor giant stars at the distance to the Sangarius stream. Of these, the {\it green filled circles} are three sources whose proper motions in both R.A. (this panel) and Dec. (lower panel) are consistent within 2 standard deviations with the expected motions in the Sangarius stream for either prograde or retrograde orbits \citep{GRILLMAIR17}. {\it (b) Lower panel:} same as {\it (a)} but for JSP proper motions in Dec., also as a function of Dec. }
\label{fig:expected_sangarius_proper_motions}
\end{figure}

Figure \ref{fig:coordinates_all_empFS_and_sangarius} also shows the positions of 19 stars from our sample, shown as {\it blue and green filled circles}, which were identified through the use of a color-magnitude diagram for metal-poor stars at the distance to the stream. In order to identify these 19 stars, we first used the DSEP \citep{DARTMOUTH} metal-poor isochrone described above. We then assumed that all 10,818 empirical faint stars in our sample were at a distance of 20 kpc and computed their absolute AB mag in the SExtractor F814W bandpass from the corresponding distance modulus. In the resulting $r-i^{+}$ color-SExtractor F814W absolute mag diagram of these sources, there were 19 stars whose absolute magnitude was within 1-mag of the above DSEP \citep{DARTMOUTH} isochrone for giant stars. The relative uncertainty in color $r-i^{+}$ was constrained to be less than 50\% before selecting the above sources. Figure \ref{fig:coordinates_all_empFS_and_sangarius} shows that the 19 sources from the Sangarius-like metal poor color-magnitude diagram and the above 2 giant stars (shown as {\it red filled circles}) fall in narrow bands in the COSMOS field, along its western and eastern edges. It is possible that the band along the western edge of the COSMOS field consists of candidate members along the eastern edge of the Sangarius stream. The sources along the eastern boundary of the COSMOS field are interesting; \citet{GRILLMAIR17} found several more or less parallel, north-south streams in the region he termed “the Orphanage”. However, the streams Scamander and PS1-D in this region are offset by $\sim 8\arcdeg$ from Sangarius, far beyond the limits of the COSMOS field. It may be that the streams in the Orphanage are considerably more sub-structured than could be discerned from the limited resolution afforded by the much brighter magnitude limits of the SDSS and Pan-STARRs data. The band of sources at the eastern end of the COSMOS field is possibly a parallel splinter of the Sangarius stream and would add to the growing list of streams showing fibrous or banded structure, including the Anticenter Stream and the Eastern Banded Structure \citep{GRILLMAIR2006}.

Of the 21 candidates shown in Figure \ref{fig:coordinates_all_empFS_and_sangarius} as {\it filled circles}, only 3 had counterparts in Gaia DR2 with Gaia parallaxes $< 0$. We visually inspected cutouts of the 21 sources in the HSC and ACS coadds, to see if any of them were resolved galaxies or AGN, and to see if there were any edge effects that might have compromised their photometry or astrometry. All of the sources were point-like and they had no contamination or truncated edges within 10 arcsec. In addition, we searched for X-ray counterparts for these 21 sources in the Chandra-COSMOS Legacy Survey Point Source Catalog \citep{CIVANO16,MARCHESI16} but found none within 7 arcsec. Therefore we believe none of these sources are unresolved AGN.

Figure \ref{fig:coordinates_all_empFS_and_sangarius} also shows as {\it red open diamonds} the positions of two giant stars from the Gaia-ACS sample of 3,939 stars that were unconstrained in SExtractor F814W AB mag. The two stars are brighter than AB 18.5 mag and are thus not part of the sample of JSP sources because they are saturated in HSC and ACS. Their Gaia parallaxes suggest that they are at distances similar to the Sangarius stream. However, their classification as giant stars was based on solar metallicity models, which is inconsistent with the stellar population of the Sangarius stream. If the stars were metal-poor instead, they would be $\sim$36 times less luminous than giant stars. In view of these caveats, these two stars are very questionable as candidates of Sangarius.

The proper motions of the candidates in the Sangarius stream and the possible substructure or stream east of it can be compared with the expected motions in the Sangarius stream. In Figures \ref{fig:expected_sangarius_proper_motions}a and \ref{fig:expected_sangarius_proper_motions}b we plot respectively the R.A. and Dec. proper motions of all 10,818 empirical faint stars in our sample, as a function of Dec. Overplotted as {\it red filled circles with error bars} are the proper motions of the candidates listed in the first two rows of Table \ref{tab:halo_stars}. Overplotted also as {\it blue filled circles} and {\it green filled circles} are the proper motions of the candidates from the metal-poor color-magnitude diagram described above. The near horizontal lines are the expected proper motions in the Sangarius stream \citep{GRILLMAIR17}, for either prograde ({\it solid line}) or retrograde ({\it dashed line}) orbits. 

It can be seen from Figure \ref{fig:expected_sangarius_proper_motions} that the two candidate sources that were selected from their high space velocities in the Toomre diagram of Figure \ref{fig:toomre_diags} have motions consistent with those expected for prograde orbits in the Sangarius stream, within 1 standard deviation. It can also be seen that, among the 19 sources selected from the metal-poor color-magnitude diagram for the Sangarius stream, 3 sources shown as {\it green filled circles} and listed in Table \ref{tab:halo_stars} under ``Candidates From Color-Magnitude Diagram'' have motions consistent within 2 standard deviations with those in this stream. Of these 3 sources, the first and the third, as listed in Table \ref{tab:halo_stars}, have R.A. and Dec. proper motions consistent with both prograde and retrograde orbits, while the second source has proper motions consistent with a retrograde orbit. The Galactic space velocities of these three stars, at the distance of 20 kpc of the Sangarius stream, are shown in Figure \ref{fig:toomre_diags}b as {\it small red filled circles}. It can be seen that these stars have high velocities, with two of them being candidate members of the halo and one of them being a candidate of the thick disk.

\section{Conclusions} \label{sec:conclusions}

Astrometry obtained by jointly processing the HSC and ACS datasets has allowed us to measure proper motions of nearly six times more sources than Gaia DR2. We have been able to derive empirical parallaxes for these sources, using ancillary Subaru SuprimeCam photometry. The proper motions of late-type main-sequence stars at $\sim$1 kpc in the COSMOS field exhibit preferential motions (relative to the LSR) directed towards the Galactic center. We have identified candidate high velocity stars, of which 6 are at relatively near distances to us, $\sim$0.5--6 kpc, and 5 are candidate members of the $\sim$20 kpc-distant Sangarius stream in the Galactic halo. The proper motions of the 5 Sangarius stream candidates are consistent with the motions previously observed in the stream, and it is possible that they are metal-poor objects that would be consistent with the halo environment. Spectroscopy of these sources is needed to confirm the above membership. We have also possibly identified a substructure or an additional stream parallel to Sangarius, $\sim 1.8\arcdeg$ to the east of it.

\section*{Acknowledgements}

We thank Dr. Davy Kirkpatrick for very insightful discussions, and the anonymous referee for very helpful suggestions.

The Hyper Suprime-Cam (HSC) collaboration includes the astronomical communities of Japan and Taiwan, and Princeton University. The HSC instrumentation and software were developed by the National Astronomical Observatory of Japan (NAOJ), the Kavli Institute for the Physics and Mathematics of the Universe (Kavli IPMU), the University of Tokyo, the High Energy Accelerator Research Organization (KEK), the Academia Sinica Institute for Astronomy and Astrophysics in Taiwan (ASIAA), and Princeton University. Funding was contributed by the FIRST program from Japanese Cabinet Office, the Ministry of Education, Culture, Sports, Science and Technology (MEXT), the Japan Society for the Promotion of Science (JSPS),Japan Science and Technology Agency (JST), the Toray Science Foundation, NAOJ, Kavli IPMU, KEK, ASIAA, and Princeton University.  

This paper makes use of software developed for the Large Synoptic Survey Telescope. We thank the LSST Project for making their code available as free software at  http://dm.lsst.org

Based in part on data collected at the Subaru Telescope and retrieved from the HSC data archive system, which is operated by Subaru Telescope and Astronomy Data Center at National Astronomical Observatory of Japan.

This work has made use of data from the European Space Agency (ESA) mission
{\it Gaia} (\url{https://www.cosmos.esa.int/gaia}), processed by the {\it Gaia}
Data Processing and Analysis Consortium (DPAC,
\url{https://www.cosmos.esa.int/web/gaia/dpac/consortium}). Funding for the DPAC
has been provided by national institutions, in particular the institutions
participating in the {\it Gaia} Multilateral Agreement.

This research used resources of the National Energy Research
Scientific Computing Center, a DOE Office of Science User Facility
supported by the Office of Science of the U.S. Department of Energy
under Contract No. DE-AC02-05CH11231.
 
%

\vspace{5mm}
\facilities{HST(ACS), Subaru(HSC), Gaia}


\software{ 
          SExtractor \citep{SEXTRACTOR}
          }



\appendix

\begin{deluxetable*}{lllc}
\tabletypesize{\footnotesize}
\rotate
\tablecolumns{4}
\tablecaption{\sc Description of Samples of Stars in the COSMOS field Used Throughout This Paper}
\tablehead{
\multicolumn{2}{l}{\sc Number of Sources\tablenotemark{a}}   & \colhead{\sc Description of Sample} & \colhead{\sc Section in Paper} \\ }
\startdata 
2,434 & \nodata & HSC-Gaia matches in all 133 HSC patches & \ref{sec:matches} \\
\nodata & 1,514 & HSC-Gaia matches in the 63 HSC coadds common to ACS; subset of the above sample & \ref{sec:matches} \\
1,135 & \nodata & ACS-Gaia matches in the 63 ACS coadds & \ref{sec:matches} \\
3,937 & \nodata & Gaia-ACS matches, unconstrained in apparent magnitude or parallax; 453 stars common to sample of 13,009 empirically selected stars & \ref{sec:FS_wrt_Gaia} \\
\nodata & 3,386 & Subset of the above; Subaru Suprime $r-i^{+}$ color in the range $-$1.25 to 3.75 mag & \ref{sec:stellar_distances_velocities} \\
\nodata & 3,367 & Gaia-ACS matches; positive Gaia parallax; subset of the above sample & \ref{sec:stellar_distances_velocities} \\
\nodata & 2,703 & Gaia-ACS matches; positive Gaia parallaxes; Subaru Suprime $r-i^{+}$ color in the range $-$1.25 to 3.75 mag; relative phot. \\ 
& & uncertainties in $r$ and $i^{+} < 0.2$; subset of the above & \ref{sec:stellar_distances_velocities} \\
\nodata & 1,010 & HSC-ACS-Gaia matches, fainter than G of 18.5; Gaia parallax $>$ 0; subset of the above sample of 3,367 sources & \ref{sec:gaia_wrt_gaia} \\
\nodata & 972 & HSC-ACS-Gaia matches, fainter than G of 18.5; Gaia parallax $>$ 0; Gaia R.A. proper motions in the range $-$30 to $+$20 mas/yr. \\
& & Subset of the above sample & \ref{sec:gaia_wrt_gaia} \\
\nodata & 925 & HSC-ACS-Gaia matches, fainter than G of 18.5; Gaia parallax $>$ 0; Gaia Dec. proper motions in the range $-$20 to $+$10 mas/yr. \\
& & Subset of the above sample of 1,010 stars & \ref{sec:gaia_wrt_gaia} \\
\nodata & 148 & Subset of the above sample of 1,010 Gaia stars; Subaru Suprime $r-i^{+}$ colors \\
& & in the range 2.75 to 3.75 for late-type; main-sequence luminosity class & \ref{sec:stellar_distances_velocities} \\
13,009 & \nodata & Empirically-selected stars in ACS coadds of SExtract F814W AB mag 19--26; includes 453 Gaia stars in the above sample of \\
& & 3,937 sources & \ref{sec:FS_wrt_Gaia} \\
\nodata & 11,519 & Empirically-selected stars, HSC matched to ACS; SExtract F814W AB mag in the range 19--25; excludes 264 Gaia stars \\
& & that were common to the above sample of 1,010 stars; subset of the above &  \ref{sec:stellar_distances_velocities} \\
\nodata & 10,818 & Subset of the above; Subaru Suprime $r-i^{+}$ colors in the range $-$1.25 to 3.75 mag relative phot. uncertainties \\ 
& & in $r$ and $i^{+} < 0.2$ & \ref{sec:stellar_distances_velocities} \\
\nodata & 10,816 & Subset of the above; empirically-derived main-sequence luminosity class & \ref{sec:stellar_distances_velocities} \\
\nodata & 8,358 & Subset of the above; late-type main sequence stars with Subaru Suprime $r-i^{+}$ colors in the range 2.75 to 3.75 mag &  \ref{sec:stellar_distances_velocities} \\
\nodata & 1,540 & Subset of the above; late-type main-sequence stars with SExtract F814W AB mag in the range 24 to 25 & \ref{sec:stellar_distances_velocities} \\
12,529 & \nodata & Combination of the above samples of 11,519 empirically selected stars and 1,010 Gaia stars & \ref{subsec:pseudo_pm} \\
\enddata
\tablenotetext{a}{Number of sources is listed in the second column of the table if the sample is a subset of another sample.}
\label{tab:samples_description}
\end{deluxetable*}

\section{Description of Stellar Samples Used Throughout This Paper} \label{sec:appendix_A}

Various samples of stars, such as from Gaia, HSC, and ACS are used throughout this paper. In order to clarify these samples, Table \ref{tab:samples_description} briefly describes them and lists the number of stars contained in them and the initial Section where they are mentioned.

\section{Parallax and Total Proper Motion Relations for High-Latitude Gaia DR2 Sources Distributed At Various Longitudes}
\label{sec:appendix_B}

Section \ref{sec:stellar_distances_velocities} presented polynomial fits to the relations between Gaia DR2 parallax and total proper motion in samples of constrained $r - i^{+}$ color and luminosity class. A limitation of the relations in Table \ref{tab:group_color_fits_Gaia} is that the total motion consists of only total proper motion or tangential velocity and does not include radial velocity. In order to assess the validity of our relations, we obtained the total proper motions and parallaxes of all Gaia DR2 stars with Galactic latitude $b > 40 \arcdeg$ and distributed at all Galactic longitudes. In this way, unknown radial velocities could be assumed to randomly span most possible values and the extinction could be taken as low as that in the COSMOS field. As in Section \ref{sec:matches}, only single, non-variable Gaia DR2 sources fainter than $G$ 18.5 mag and with parallax $>0$ were considered. The spectral type of sources was estimated from the Gaia color $G_{BP} - G_{RP}$ in the Vega system, after first generating synthetic $G_{BP} - G_{RP}$ colors of main-sequence and giant stars by convolving empirical templates of stellar spectra \citep{KESSELI2017} with the transmission functions of the $G_{BP}$ and $G_{RP}$ filters \citep{GAIA-MISSION,GAIA-DR2}. The luminosity class (main-sequence or giant) of each Gaia DR2 source was discriminated by comparing $G$ with the apparent magnitude derived from the distance modulus for either main-sequence or giant stars, similarly to Section \ref{sec:stellar_distances_velocities}.

These high-latitude Gaia DR2 stars were binned in representative groups, for comparison of their parallax vs. total proper motion relation to those groups in Table \ref{tab:group_color_fits_Gaia} of approximately similar spectral type and luminosity class ranges. The data sets from which the groups were drawn consisted of 729,514 stars with $2.75 < G_{BP} - G_{RP} < 3.75$ or spectral type M5--M8, 3'810,892 stars with $0.8 < G_{BP} - G_{RP} < 1.2$ or spectral type G4-K3, and 2'470,402 stars with $1.2 < G_{BP} - G_{RP} < 1.5$ or spectral type K4--K7. Table \ref{tab:group_color_fits_all_high_lat_Gaia} lists the representative bins of main-sequence and giant stars drawn from the above three data sets. These representative bins are each approximate supersets of the respective bins listed in Table \ref{tab:group_color_fits_Gaia}.

\begin{deluxetable*}{cccccc}
\tabletypesize{\footnotesize}
\rotate
\tablewidth{6.5in}
\tablecolumns{6}
\tablecaption{\sc Polynomial Fits To the Relation of Gaia DR2 Parallax As A Function of Total Proper Motion, Color, and Luminosity Class For Representative Groups Drawn From All Gaia DR2 Stars at High Galactic Latitude \tablenotemark{a}}
\tablehead{
\colhead{\sc Color}   & \colhead{\sc Spectral Type} & \colhead{$a_{0}$} & \colhead{$a_{1}$} & \colhead{$a_{2}$} & \colhead{$a_{3}$} \\
\colhead{$G_{BP} - G_{RP}$} & \colhead{}                  & \colhead{}   & \colhead{}            & \colhead{} & \colhead{} 
}
\startdata 
\multicolumn{6}{l}{\sc Main Sequence Stars} \\
0.80 to  1.20     & G4 to K3 &  0.52$\pm$0.11 & $-$0.017$\pm$0.018 & 6.7$\times 10^{-4}\pm6.5\times 10^{-4}$  & 1.03$\times 10^{-5}\pm5.9\times 10^{-6}$ \\
1.20 to  1.50     & K4 to K7 &  0.592$\pm$0.016 &  0.0046$\pm$0.0014  & 1.16$\times 10^{-4}\pm 2.0\times 10^{-5}$ & \nodata \\
2.75 to  3.75     & M5 to M8 &  1.997$\pm$0.016 &  0.0225$\pm$0.0014  & 2.9$\times 10^{-4}\pm 2.0\times 10^{-5}$ & \nodata \\
\multicolumn{6}{l}{\sc Giant Stars} \\
0.80 to  1.20     & G4 to K3 & 0.01813$\pm$0.00021  & $-1.39\times 10^{-4}\pm4.5\times 10^{-5}$ & $6.2\times 10^{-6}\pm 1.9\times 10^{-6}$ & \nodata \\
1.20 to  1.50     & K4 to K7 & 0.014929$\pm$0.00053  & 2.8$\times 10^{-5}\pm 2.4\times 10^{-5}$  & \nodata & \nodata \\
2.75 to  3.75     & M5 to M8 & 0.02198$\pm$0.00038  & $8.68\times 10^{-5}\pm 3.72\times 10^{-5}$             & \nodata & \nodata \\
\enddata
\tablenotetext{a}{Polynomial fits to the relation of Gaia DR2 parallax as a function of total proper motion, also as a function of Gaia DR2 color $G_{BP}-G_{RP}$ and luminosity class (main-sequence or giant stars). These relations were established from 729,514 stars with $2.75 < G_{BP} - G_{RP} < 3.75$, 3'810,892 stars with $0.8 < G_{BP} - G_{RP} < 1.2$, and 2'470,402 stars with $1.2 < G_{BP} - G_{RP} < 1.5$, and all with positive Gaia DR2 parallax. The polynomial fits are of the form $\pi = a_{0} + a_{1}\mu_{total} + a_{2}\mu_{total}^{2} + a_{3}\mu_{total}^{3}$ where it is implied that $a_{2}$, $a_{3}$ are 0 if not listed. The quoted errors in the polynomial fit coefficients are $\pm$ 1 standard deviation. }
\label{tab:group_color_fits_all_high_lat_Gaia}
\end{deluxetable*}

\begin{figure}[htb]
\hspace{-0.5in}
\includegraphics[angle=0,width=4.0in]{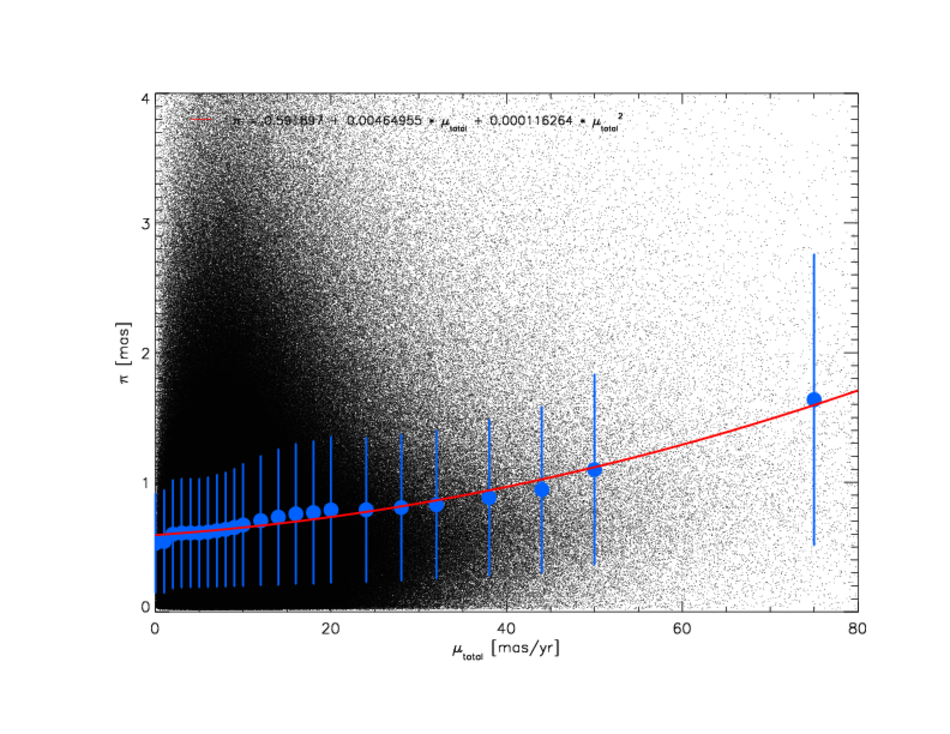}
\caption{Gaia DR2 catalog parallax as a function of total proper motion for late-K-type main sequence Gaia DR2 stars (whose luminosity class and spectral type were determined as described in the text), with Gaia colors in the interval $1.2 < G_{BP} - G_{RP} < 1.5$, and with parallaxes $>$ 0 indicating valid Gaia DR2 solutions. These sources were drawn from 2,470,402 stars at Galactic latitude $b > 40 \arcdeg$ and distributed at all Galactic longitudes. The {\it small black symbols} correspond to individual stars. The {\it large blue symbols with error bars} are the proper-motion-binned means and standard deviations of parallax. The {\it solid red line} is a second order polynomial fit to the proper-motion-binned means.}
\label{fig:pi_vs_mu_M_dwarfs_all_highlat}
\end{figure}

Upon comparing Tables \ref{tab:group_color_fits_all_high_lat_Gaia} and \ref{tab:group_color_fits_Gaia} it can be seen that the Gaia DR2 parallax vs. proper motion relations of all high-latitude stars are ``shallower'' (have a smaller slope) than those of the COSMOS field. Figure \ref{fig:pi_vs_mu_M_dwarfs_all_highlat} shows this relation for K4-K7 Gaia stars at all high latitudes, for comparison with Figure \ref{fig:pi_vs_mu_M_dwarfs} (M2-M3 Gaia stars in COSMOS). Figures \ref{fig:pi_vs_mu_M_dwarfs} and \ref{fig:pi_vs_mu_M_dwarfs_all_highlat} and Tables \ref{tab:group_color_fits_Gaia} and \ref{tab:group_color_fits_all_high_lat_Gaia} show that the relation for all high-latitude stars is a factor of $\sim$2 shallower than that for the COSMOS field. The ``zero-intercept'' or the 0th-order coefficient of the fits are comparable to within $\sim$20\%, which represent the most distant stars in these relations. We believe that the above comparison gives an idea of the uncertainty of these relations in the absence of radial velocity information.

\begin{figure*}[htb]
\centering
\includegraphics[angle=0, width=5.7in]{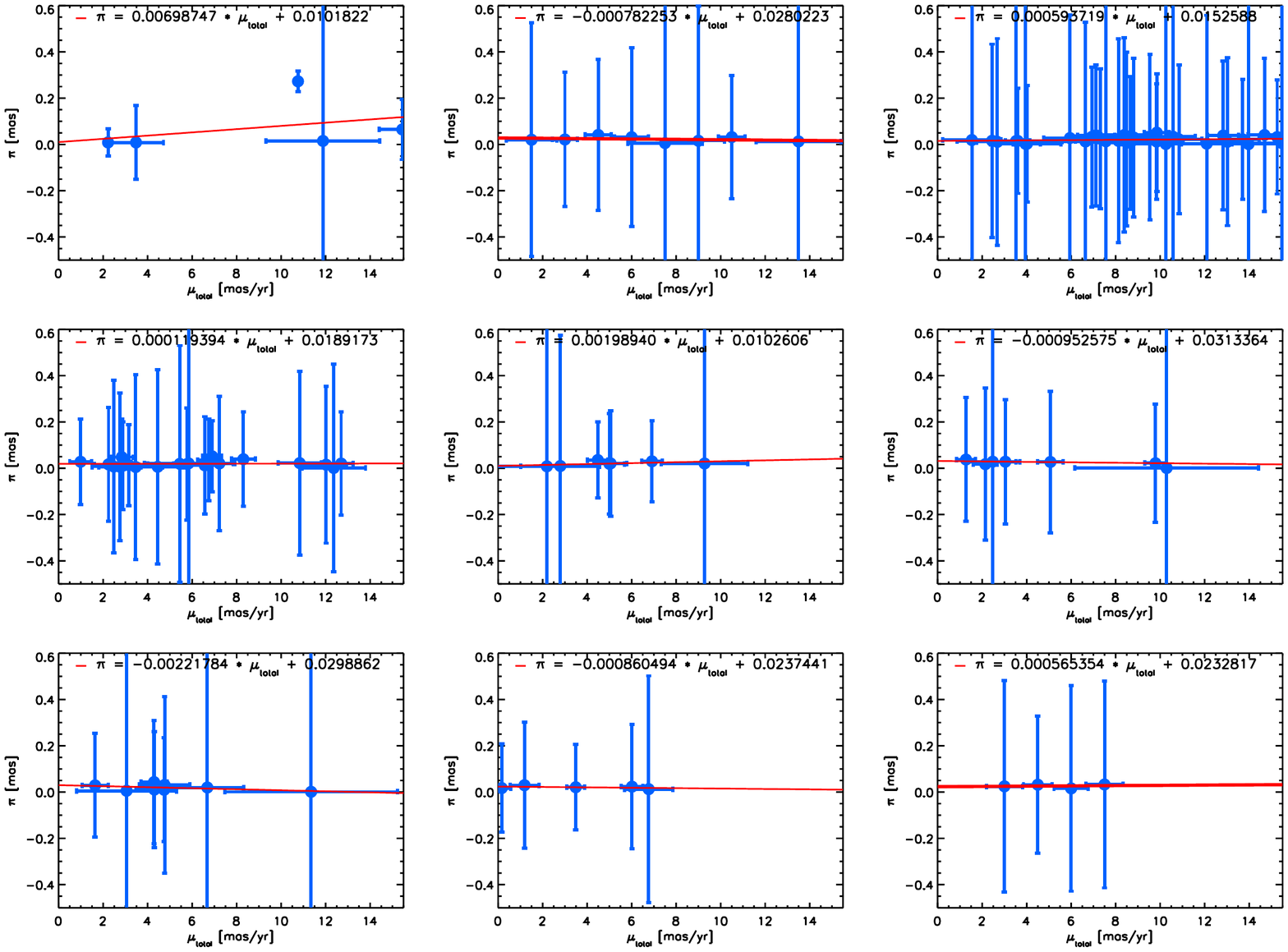}
\caption{Gaia DR2 catalog parallax as a function of total proper motion for giant stars of type K8-M1 (whose luminosity class and spectral type were determined as described in the text) and with parallaxes $>$ 0 indicating valid Gaia DR2 solutions. The {\it upper-left panel} is for the 5 ACS-HSC-Gaia DR2 giant stars in the COSMOS field, identified from the sample of stars with Subaru SuprimeCam colors in the interval $0.5 < r - i^{+} < 0.8$. The {\it remaining panels} are for Gaia DR2 stars in fields of area 1.64 square degrees each, analogous to the COSMOS field but distributed at various locations at Galactic latitude $b > 40 \arcdeg$. These stars were identified from the sample of sources with Gaia DR2 colors in the interval $1.6 < G_{BP} - G_{RP} < 2.0$). The {\it blue filled circles with error bars} correspond to individual stars, where the error bars are the Gaia DR2 catalog uncertainties. The {\it solid red lines} are linear fits to the data of stars in each field. The panels, starting from the middle top and running from left-to-right and and from top-to-bottom, are for fields at Galactic longitude, latitude ($l$,$b$) of (297,42), (357,42), (57,42), (117,42), (357,60), (117,60), and (57,$-$42), respectively, in degrees. A field at (237,60) had fewer than 3 giant stars and is omitted. }
\label{fig:pi_vs_mu_KM_giants_cosmos_and_highlat}
\end{figure*}

We also compared the relations of parallax vs. proper motion for some of the most distant stars, namely giants of spectral types K8-M1 in either the COSMOS field or in eight high-latitude fields. The latter were chosen to be 1.64 square degrees in size, analogous to the size of the COSMOS field covered by ACS, at Galactic latitudes $b > 40 \arcdeg$ (with one field at $b < -40 \arcdeg$), and equally distributed in Galactic longitude. Figure \ref{fig:pi_vs_mu_KM_giants_cosmos_and_highlat} shows the parallax vs. total proper motion relations for these fields. The fit parameters of these fields are indicated in each panel in Figure \ref{fig:pi_vs_mu_KM_giants_cosmos_and_highlat}, including for the COSMOS field giant stars in the {\it upper left panel}, and also in Table \ref{tab:group_color_fits_Gaia} for the latter.

\begin{figure}[htb]
\centering
\includegraphics[angle=0, width=3.2in]{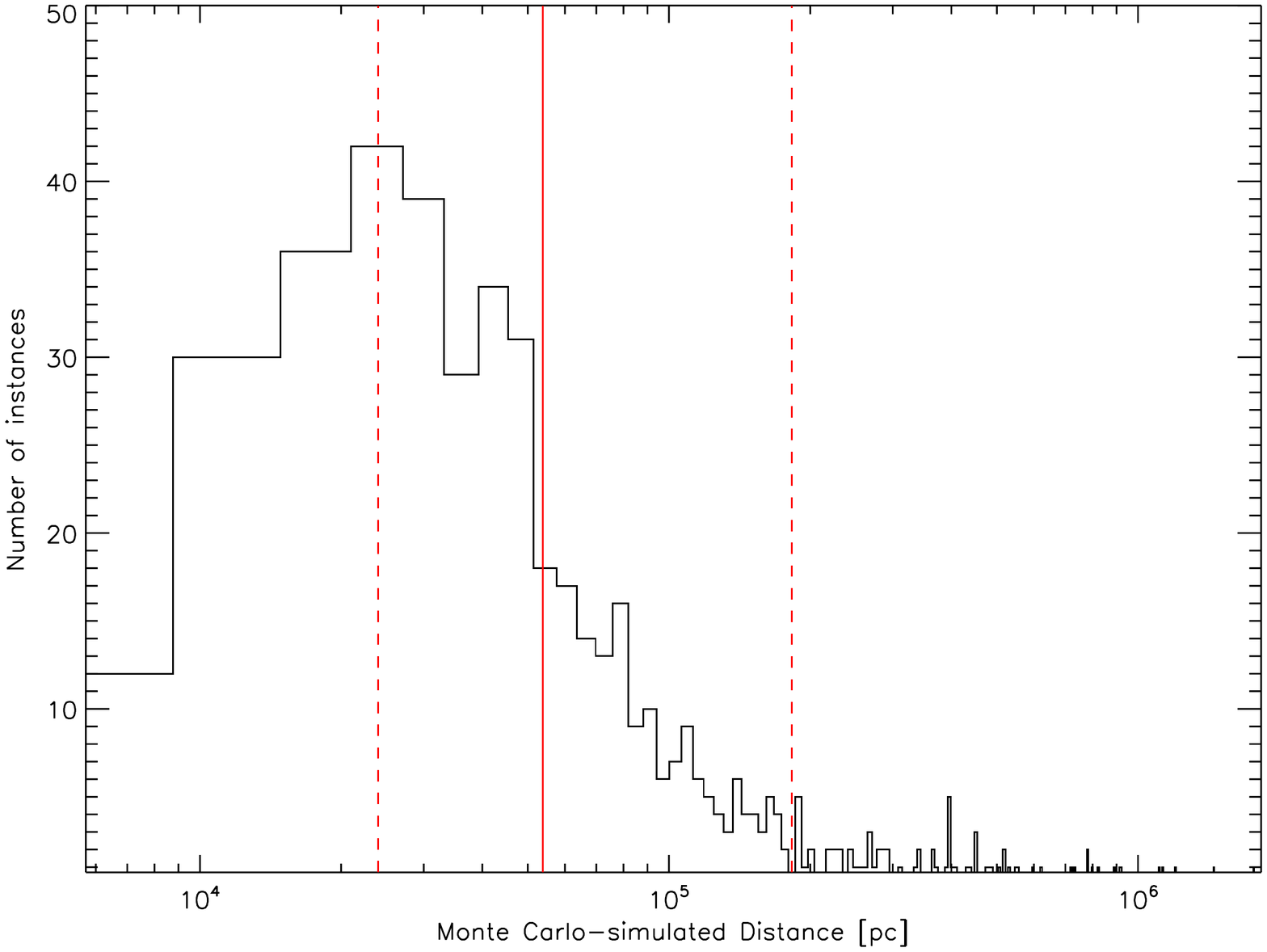}
\caption{Monte Carlo simulation of distances for the giant star of spectral type M0 listed in the first row of Table \ref{tab:halo_stars}. These distances are the result of generating 500 random Gaussian deviates to the Gaia proper motions and parallaxes of K8-M1 giant stars plotted in Figure \ref{fig:pi_vs_mu_KM_giants_cosmos_and_highlat} and to the JSP proper motions of this star (Table \ref{tab:halo_stars}). Linear fits were generated for each set of Gaia parallaxes and proper motions, of the form listed in Table \ref{tab:group_color_fits_Gaia}, and the JSP proper motions were applied in the linear fit to generate a parallax and in turn a distance. The {\it solid red vertical line} is the median distance and the {\it vertical dashed lines} represent the $\pm 66.67$\% points of the distribution, corresponding to asymmetrical $\pm 1 \sigma$ (1 standard deviations) }
\label{fig:MonteCarlo_distances_simul}
\end{figure}

Figure \ref{fig:pi_vs_mu_KM_giants_cosmos_and_highlat} shows that the Gaia DR2 parallax uncertainties are very large for these distant sources. The linear fits are mostly flat for total proper motions up to 15 mas/yr in most cases and there are generally few available sources in each field, although the fields at ($l$,$b$) of (357,42) and (57,42) have 32 and 20 sources, respectively. If radial velocities were available, the slopes of the linear fits would be smaller in absolute value because all abcissae would be incremented. Thus, the slightly negative slopes in the fields at (297,42), (117,42), (357,60) and (117,60) are most likely due to the lack of radial velocities.

The resulting fits of the relations of Gaia parallax vs. total proper motion for distant stars have very large uncertainties. For example, the linear fit parameters for the COSMOS field giants of K8 to M1-type listed in Table \ref{tab:group_color_fits_Gaia} have relative uncertainties of $\sim$380\% and 33\% in the zero-intercept ($a_{0}$) and slope ($a_{1}$), respectively. In view of the large uncertainties of parallax, the model of distance as a reciprocal of a single value of parallax fails and distance must be assessed statistically instead. For this purpose, we re-evaluated the distance of the two giant stars listed in the first two rows of Table \ref{tab:halo_stars}. Given the uncertainties of Gaia DR2 proper motion and parallax, and the uncertainties of JSP proper motion of these two sources listed in Table \ref{tab:halo_stars}, we generated Gaussian deviations for these three parameters and calculated linear fits similar to that for K8-M1 giants in Table \ref{tab:group_color_fits_Gaia} and the upper-left panel in Figure \ref{fig:pi_vs_mu_KM_giants_cosmos_and_highlat}. Gaussian deviations that yielded JSP parallax $< 0$ were discarded and the Monte Carlo simulations were continued until 500 successful simulations were achieved. Figure \ref{fig:MonteCarlo_distances_simul} shows the histogram of the distances that were computed for the M0 III star. The median distance of the 500 simulations was taken as the distance to the source. The asymmetric lower and upper standard deviations were found at the distance values where the histogram contained 67\% of the points below and above the median, respectively. These values are listed in the last column of Table \ref{tab:halo_stars} for these two stars.

\section{High Velocity Stars Drawn From the Sample of Gaia-ACS Matches Distinct From Empirically Selected Stars} \label{sec:appendix_C}

\begin{deluxetable*}{ccccccccc}
\tabletypesize{\footnotesize}
\rotate
\tablewidth{6.5in}
\tablecolumns{9}
\tablecaption{\sc Candidate high-velocity Stars In the Gaia DR2 Proper Motion Sample}
\tablehead{
\colhead{\sc R.A. (J2000.0)} & \colhead{\sc Dec. (J2000)} & \multicolumn{2}{c}{\sc Gaia DR2 Proper Motions\tablenotemark{a}} & \colhead{\sc ACS F814W mag} & \colhead{\sc Color $r - i^{+}$\tablenotemark{b}} & \colhead{\sc Spectral Type\tablenotemark{c}} & \colhead{\sc Luminosity Class\tablenotemark{c}} & \colhead{\sc Distance\tablenotemark{d}}\\
\colhead{} & \colhead{} & \colhead{$\mu_{\alpha}$} & \colhead{$\mu_{\delta}$} & \colhead{} & \colhead{} & \colhead{} & \colhead{} & \colhead{} \\
\colhead{h, m, s} & \colhead{d, m, s} & \colhead{mas/yr} & \colhead{mas/yr} & \colhead{mag [AB]} & \colhead{mag [AB]} & \colhead{} & \colhead{} & \colhead{kpc} }
\startdata 
09 57 59.04502 & $+$02 42 39.6025 &  $-$20.70$\pm$0.53 & $-$1.54$\pm$0.38 & 18.800 & $+$2.827$\pm$0.013 & mid-to-late M & III & 2.43$^{+3.7}_{-1.3}$ \\
10 02 31.98050 & $+$02 44 41.15652 &  $+$22.59$\pm$0.81 & $-$45.58$\pm$0.81 & 18.855 & $-$0.3918$\pm$0.0042   & O8 & V & 1.27$^{+1.58}_{-0.57}$ \\
10 01 11.28869 & $+$02 47 42.95436 &  $-$91.72$\pm$0.71 & $-$23.44$\pm$0.75 & 18.856 & $-$0.0542$\pm$0.0038   & A8 & V & 0.93$^{+0.65}_{-0.30}$ \\
10 00 24.46502 & $+$01 52 48.3867 &  $-$16.81$\pm$0.52 & $-$37.665$\pm$0.56 & 18.874 & $-$0.4052$\pm$0.0036 & O4 & V & 1.96$^{+3.2}_{-0.92}$ \\
09 59 11.15546 & $+$02 00 46.6052 &  $+$14.86$\pm$0.83 & $-$36.17$\pm$0.72 & 19.040 & $+$0.4832$\pm$0.0037 & K7 & V & 2.0$^{+3.9}_{-1.1}$ \\
09 58 15.76378 & $+$02 05 17.0138 &  $-$62.86$\pm$1.02 & $-$28.17$\pm$1.02 & 19.236 & $+$0.1366$\pm$0.0052 & G3 & V & 0.76$^{+1.25}_{-0.28}$ \\
\enddata
\tablenotetext{a}{Proper motions are from the Gaia DR2 catalog, corrected for the effect of Solar motion, as explained in Section \ref{sec:stellar_distances_velocities}.}
\tablenotetext{b}{Colors are from the Subaru Suprime-Cam COSMOS2015 survey \citep{COSMOS2015}.}
\tablenotetext{c}{Spectral types were derived from synthetic colors of main sequence and giant stars using models from \citet{KESSELI2017}, as explained in Section \ref{sec:stellar_distances_velocities}. Luminosity class was inferred from the observed SExtract F814W AB magnitude and absolute magnitudes of main sequence and giant stars (see references in Section \ref{sec:stellar_distances_velocities}). }
\tablenotetext{d}{Sources are from the sample of 1,010 Gaia-ACS matched stars with Subaru Suprime $r-i^{+}$ colors in the range $-$1.25 to 3.75, SExtract F814W AB mag $>$ 18.5, and positive Gaia parallaxes. In view of large relative uncertainties of Gaia DR2 parallaxes for these sources, Monte Carlo simulations of parallax were carried out to derive asymmetric distributions of simulated distance, analogous to the simulations in Appendix \ref{sec:appendix_B}; see also \citet{GAIADR2PARALLAX}. }
\label{tab:gaia_high_vel}
\end{deluxetable*}

Section \ref{sec:FS_wrt_Gaia} indicated that our sample of stars with well-determined JSP proper motions, Subaru Suprime $r-i^{+}$ colors in the range -1.25 to 3.75 mag, and SExtract F814W AB mag fainter than 18.5, consisted of 1,010 Gaia DR2 sources matched to ACS, and 10,818 empirically selected stars. The color range was established to exclude any abnormal stars, and the magnitude limit was meant to exclude saturated sources. In the sample of Gaia stars, which is distinct from the above sample of empirically selected stars, and using the techniques described in Section \ref{sec:stellar_distances_velocities}, six stars were identified as having high Galactic space velocities (shown as statistical ranges of velocities by the {\it dashed green lines} in Figure \ref{fig:toomre_diags}). In Table \ref{tab:gaia_high_vel} we list observational data of these sources. These sources are inconsistent with membership in the Sangarius stream, which is at 20 kpc from us. However, we list the sources as reference for any further observations. These stars are high velocity candidates, for which halo membership determination requires further follow up, such as measurements of their metallicity. Alternatively, their high velocities might instead be due to gravitational encounters.



\bibliographystyle{aasjournal}
\bibliography{bibli}

\end{document}